\documentclass[lettersize,journal]{IEEEtran}
\pdfoutput=1
\usepackage{amsmath,amsfonts}
\usepackage{algorithmic}
\usepackage{algorithm}
\usepackage{array}
\usepackage[caption=false,font=normalsize,labelfont=sf,textfont=sf]{subfig}
\usepackage{textcomp}
\usepackage{stfloats}
\usepackage{url}
\usepackage{verbatim}
\usepackage{graphicx}
\usepackage{cite}
\usepackage{enumitem}
\usepackage[acronym]{glossaries}
\usepackage{cleveref}
\usepackage{bm}

\newenvironment{talign}{\equation\aligned}{\endaligned\endequation}

\crefname{figure}{Fig.}{Figs.}
\crefname{equation}{}{}

\begin{document}

\title{Generalized Passivity Sensitivity Methodology\\for Small-Signal Stability Analysis}

\author{Dongyeong Lee, ~Francisco Javier Cifuentes Garcia, and Jef Beerten
\vspace{-0.5cm}

\thanks{This work was supported by the DIRECTIONS Project through the Energy Transition Fund, FOD Economy, Belgium.}

\thanks{Dongyeong Lee, Francisco Javier Cifuentes Garcia, and Jef Beerten are with Department of Electrical Engineering (ESAT\,-\,ELECTA), Katholieke Universiteit Leuven, Leuven 3000, Belgium, and also with EnergyVille, Genk 8310, Belgium (e-mail: dongyeong.lee@kuleuven.be).}}


\maketitle

\begin{abstract}
This paper proposes a generalized passivity sensitivity analysis for power system stability studies. 
The method uncovers the most effective instability mitigation actions for both device-level and system-level investigations. 
The particular structure of the admittance and nodal models is exploited in the detailed derivation of the passivity sensitivity expressions.
These proposed sensitivities are validated for different parameters at device-level and at system-level.
Compared to previous stability and sensitivity methods, it does not require detailed system information, such as exact system eigenvalues, while it provides valuable information for a less conservative stable system design. 
In addition, we demonstrate how to utilize the proposed method through case studies with different converter controls and system-wide insights showing its general applicability.
\end{abstract}



\section{Introduction}
\IEEEPARstart{P}{}ower systems are currently undergoing a major change in their dynamic characteristics as more renewable energy resources are integrated via power electronics (PE) converters \cite{IEEEstability2020}. 
In addition, High Voltage Direct Current (HVDC) systems and Flexible Alternating Current Transmission System (FACTS) devices based on Voltage Source Converters (VSC) are increasingly installed around the world to benefit from their high controllability and system support capabilities \cite{HVDCoverview}. 

While modern PE-based devices offer higher degree of controllability at different time-scales compared to Synchronous Generators (SG), their behavior is predominantly dictated by their control systems which are undisclosed Intellectual Property (IP) of the manufacturers. 
Therefore, to guarantee safe integration of PE-based devices, small-signal stability studies are carried out in the Frequency-Domain (FD) where the PE converters can be represented by their input/output characteristics thus preserving the vendor's IP rights \cite{ZvsSS, Sun2022}. 
Furthermore, negative contribution to system stability from the components and their controls can be determined by means of sequential FD eigenvalue sensitivity calculations \cite{Grey-Box, On-Relationship}. 

However, FD eigenvalue sensitivity calculations relating impedance models to the changes in critical state-space modes are not accurate in the frequency-domain ($j\omega$), but in the $s$-domain with $s=\sigma \pm j\omega$, and thus only applicable to black-box models for oscillatory modes with $|\sigma| \ll \omega$. 
This assumption is also at the core of the positive mode damping (PMD) stability criterion, which is a generalization of the positive net damping (PND) criterion \cite{Sainz17, Cheah17}, proposed in \cite{PMD2021} and further demonstrated in \cite{Orellana2023} for system-wide high-frequency stability analysis.

Alternatively, to overcome said assumptions while being black-box compatible, an analytical model can be obtained from the frequency response data via vector fitting algorithms at the expense of higher computational burden and model order \cite{VF}, which might hinder their application to large-scale systems.
In addition, the required eigenvalue information to perform FD eigenvalue sensitivity analysis may not be available due to the computational burden for large-scale systems that comes from finding roots of its determinant from the entire system matrix.
Instead, the application of the Generalized Nyquist Criterion (GNC) assuming standalone-stable subsystems reduces to a simple no-encirclement check of the critical point $(-1,j0)$ by the eigenvalue loci of the minor-loop matrix \cite{Maciejowski1989}. Since the proximity of a given locus to the critical point is a measure of the stability margin, the authors in \cite{MTHVDCstability} have recently demonstrated in a four-terminal HVDC system that calculating the sensitivity of the critical locus with respect to the HVDC converter impedance and control parameters can bring meaningful root-cause and mitigation insights. 
This method only requires frequency-domain information and thus overcomes the downsides of the aforementioned approaches. 
However, mitigation based on the sensitivity of a specific critical eigenvalue locus to a control parameter or device's impedance cannot always guarantee stability improvements. 
The reason is that the critical locus can shift to another loop-gain matrix eigenvalue due to the simultaneous change in all eigenvalues after any control or system update. 

To tackle this challenge, the passivity theorem has been extensively used to guarantee closed-loop stability considering standalone-stability and additional subsystems conditions \cite{bao2007}, finding many applications in power \cite{Dey2023,Chatterjee23,Wang2024,Josep2024}. The passivity theorem states that if two standalone-stable passive subsystems are interconnected in a feedback manner, the resulting closed-loop system is stable and passive \cite{bao2007}. 
The theorem's conservatism can be reduced by considering the passivity contribution of both subsystems to guarantee closed-loop passivity frequency-wise so oscillatory instabilities do not occur in specific frequency ranges \cite{bao2007}. 
This is an advantageous property since achieving passivity across the entire frequency range is not realistic for most power system assets such as SG-based power plants and VSC-based devices \cite{Dey2021}, especially at low frequencies \cite{Feifan2024}. While the passivity theorem provides no guarantees outside of the passive ranges, extending the frequency range of passivity effectively mitigates any risk of instabilities in said range. 
Therefore, it can be used to design improved controls and define robust mitigation measures. 

In this paper, we overcome the limitations of previous works by proposing a passivity sensitivity method and validating its capabilities to provide clear insights for a stable system design. 
Other passivity-based methods are often limited to simple analytical models and/or by numerous iterations involving multiple parameters changes to understand their effect on the device-level passivity characteristics. 
However, the proposed method of passivity sensitivity allows for a more efficient methodical approach, valuable for both device-level control design and system-level analysis.
This procedure is compatible with black-box models because it uses only frequency-domain data without any state-space eigenvalues information, contrary to the FD eigenvalue sensitivity analysis \cite{Grey-Box}.
Furthermore, it provides not only the magnitude but also directional information, which are essential to guide instability-mitigation actions.
The mathematical derivation of parametric and system-level passivity sensitivities, is validated with application cases demonstrating both device-level and system-level design guidelines. 
The results show that the proposed method can be a useful design tool for power systems and connected devices as it presents less conservatism while ensuring passivity and stability for a broader time-scale range.

The remainder of the paper is structured as follows. Section II presents a literature review to recap the limitations of currently available stability analysis methods. 
Section III introduces the derivation and validation of the proposed parametric passivity sensitivity via an illustrative case study involving a single converter. 
The method is further extended for system-level studies and tested with an additional application case.
Section IV demonstrates the applicability of the proposed approach to general power systems by introducing a more realistic case study involving several PE converters, SGs and a series-compensated line. 
The analysis results of each case study are verified via PSCAD/EMTDC simulations.
Section V summarizes the main findings of this paper.

\section{Review and Limitations of Frequency Domain \\Stability Analysis}
\begin{figure}
    \centering
    \includegraphics[width=0.9\linewidth]{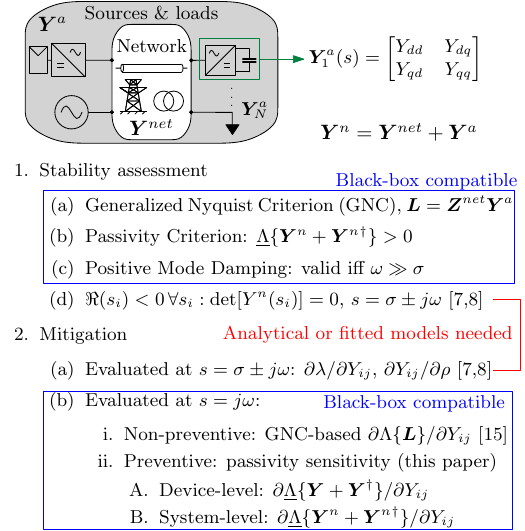}
    \caption{Summary of FD analysis methods.}
    \label{fig:summary}
    \vspace{-0.5cm}
\end{figure}
This section provides a review of existing stability and sensitivity analysis methods.
It reveals the research gaps and highlights the needs for a novel analysis method. \cref{fig:summary} shows a summary of FD stability-related analysis methods.
\subsection{Review of Frequency Domain System Model}
This subsection describes the commonly used FD system model for power systems stability studies \cite{On-Relationship, Root-Cause}, which is based on the following admittance-based description
\begin{equation}
\begin{cases}
    \boldsymbol{i}^n(s) = \underbrace{[\boldsymbol{Y^{net}}(s) + \boldsymbol{Y^{a}}(s)]}_{\boldsymbol{Y^{n}}} \boldsymbol{v}^n(s) \\
     \boldsymbol{v}^n(s) = \underbrace{[\boldsymbol{I}+\overbrace{\boldsymbol{Z^{net}}(s)\boldsymbol{Y^a}(s)}^{\boldsymbol{L}}]^{-1}\boldsymbol{Z^{net}}(s)}_{\boldsymbol{Z^{cl}}}\boldsymbol{i}^n(s)
\end{cases}    
\end{equation}
\begin{align}
&\boldsymbol{Y^n}=\boldsymbol{\mathcal{I}_i^T}\boldsymbol{Y^{b}}\boldsymbol{\mathcal{I}_i}+\boldsymbol{Y^a}+\boldsymbol{Y^c}=\boldsymbol{Y^{net}} + \boldsymbol{Y^{a}}\\
&\boldsymbol{Y^{net}}=\boldsymbol{Y^{e}}+\boldsymbol{Y^{c}} \quad\&\quad \boldsymbol{Y^{a}}=\text{diag} (\boldsymbol{Y^a}_{1}, \cdots, \boldsymbol{Y^{a}}_N)\\
&\boldsymbol{Y^{e}}=\boldsymbol{\mathcal{I}_i^T}\boldsymbol{Y^{b}}\boldsymbol{\mathcal{I}_i}\,\,\&\,\,\boldsymbol{Y^{sh}}=\boldsymbol{Y^a}+\boldsymbol{Y^c}
\end{align}
where superscript $n$, $net$, $sh$, $a$, $b$, $c$, and $e$ denote nodal, the system network, shunt components of nodal, active components, branches of the network, shunt components of the network, and edge admittance, respectively. 
The bold symbols denote vectors and matrices.
Here, active components denote non-passive devices such as VSC, SG, etc.
For a network with $N$ nodes and $B$ branches, its nodal voltages and currents are denoted by $\boldsymbol{v^n}(s) = [\boldsymbol{v}_1, \cdots, \boldsymbol{v}_N]^T \in \mathbb{C}^{2N\times 2N}$ and $\boldsymbol{i^n}(s) = [\boldsymbol{i}_1, \cdots, \boldsymbol{i}_N]^T \in \mathbb{C}^{2N\times 2N}$, respectively.
$\boldsymbol{I}\in\mathbb{R}^{2N\times 2N}$ is $2N$-dimensional identity matrix. 
Additionally, $\boldsymbol{\mathcal{I}_i}\in\mathbb{R}^{2B\times 2N}$ represents the incidence matrix containing system topology information.
$\boldsymbol{Y^b}$\,=\,$\text{diag}(\boldsymbol{Y^b}_1,\cdots,\boldsymbol{Y^b}_B)\in\mathbb{C}^{2B\times 2B}$, $\boldsymbol{Y^e}\in\mathbb{C}^{2N\times 2N}$, and $\boldsymbol{Y^c}$\,=\,$\text{diag}(\boldsymbol{Y^c}_1,\cdots,\boldsymbol{Y^c}_N)\in\mathbb{C}^{2N\times 2N}$ are branch admittance model, edge admittance model, and network shunt admittance model, respectively.
The methods for representing systems using $\boldsymbol{Y^n}(s)$ and $\boldsymbol{Z^{cl}}(s)$ are referred to as the nodal admittance model and the closed-loop impedance model, respectively, where $\boldsymbol{L}$ denotes the system's loop-gain. 
These two FD system models are equivalent as one is the inverse of the other \cite{Root-Cause}.
In this paper, the nodal admittance model is used as the formulation allows to more easily emphasize the passivity-related aspects.
\subsection{Review of Small-Signal Stability Analysis Methods}
Conventionally, power system small-signal stability has been assessed using the system's state-space (SS) model \cite{Kundur}.
\begin{equation}
     \boldsymbol{\dot{x}}(t)=\boldsymbol{A}\boldsymbol{x}(t) + \boldsymbol{B}\boldsymbol{u}(t) 
\end{equation}
where $\boldsymbol{x}(t)=[x_{1},\cdots,x_{K}]^T\in\mathbb{R}^K$ are the system's state variables and $K$ denotes the number of state variables in the system. 
The eigenvalues of the system matrix $\boldsymbol{A}$, $\boldsymbol{\lambda}=\text{eig}(\boldsymbol{A})=[\lambda_{1},\cdots,\lambda_{K}]\in\mathbb{C}^K$, describe the system's dynamic behavior. 
Specifically, if the real part is negative, $\Re\{\lambda\}<0,\,\,\, \forall\,\lambda$, the system is asymptotically stable. 
This is the state-space criterion to determine system stability.

This interpretation of system dynamics for stability can also be found through the FD models of (1), i.e.,
$\boldsymbol{\lambda}$ can be calculated from the zeros of $\text{det}[\boldsymbol{Y^n}(s)]$ or the poles of $\boldsymbol{Z^{cl}}(s)$. 
However, calculating eigenvalues through $\boldsymbol{Y^n}$ can become computationally impossible due to its complexity, which comes from finding the roots of a large matrix's determinant, especially for large systems. 

Another popular FD stability analysis method is rooted in the GNC to determine the closed-loop stability by analyzing the impedance ratio or minor-loop matrix $\boldsymbol{L}(s)$ obtained by partitioning the system into two independently stable groups \cite{Maciejowski1989, MTHVDCstability}. 
According to the GNC, if the eigenloci of $\boldsymbol{L}$ encircles (-1,$j$0) from its Nyquist plot, the interconnected system is unstable. 
However, the system partitioning points are critical and wrongly selecting them can lead to inaccurate stability conclusions. 
To address this issue, the PMD method was proposed in \cite{PMD2021, Orellana2023}. In addition to stability prediction, this method can provide the participating buses on any given oscillatory mode. 
The PMD only uses frequency information, however, it is only applicable for modes where $\omega\gg\sigma$ or for very poor damped (those with very low $\Re\{\lambda\}$). 

Besides these stability assessment methods, the passivity theorem has been widely used in control design to ensure stability within specific frequency range, based on the assumption that most power system network elements such as transmission lines, shunt capacitors/reactors, transformers, etc. are inherently passive.
This method is valuable given the unpredictability of network behavior, e.g., topology changes and contingencies that might lead to changes in the passive resonance frequencies. 
Formally, a system with a set of inputs $\boldsymbol{u}(t)$ and a set of outputs $\boldsymbol{y}(t)$ and its storage function $\boldsymbol{S}$ \cite{Dey2023} is said to be passive if \eqref{eq: Passivity Eqns}.
\begin{equation}
    \boldsymbol{u}^T(t)\boldsymbol{y}(t) \geq \frac{d\boldsymbol{S}}{dt}, \,\,\,\,\&\,\,\,\ 
    \forall\, \boldsymbol{u}(t), \boldsymbol{y}(t),\,\, \text{and}\,\, t \geq 0.
    \label{eq: Passivity Eqns}
\end{equation}
Equivalently, given a small-signal frequency-domain model as a $N$-dimensional transfer matrix $\boldsymbol{y(s)} = \boldsymbol{G(s)}\boldsymbol{u(s)}$, the passivity conditions can be stated as follows \cite{bao2007}
\begin{enumerate}
    \item $\boldsymbol{G}(j\omega)$ has no right-half plane (RHP) poles, i.e., stable.
    \item $\underline{\Lambda}\{\boldsymbol{G}(j\omega) +\boldsymbol{G}^\dagger(j\omega)$\} \,$\geq$ \,0,\,\,\,\,$\forall$$\omega$ $\in [-\infty, \infty]$
\end{enumerate}
where superscript $\dagger$ denotes the Hermitian operator of the matrix, and $\underline{\Lambda}$ denotes the minimum 
 of a given sorted-eigenvalue set $\Lambda$, i.e. $\underline{\Lambda}=\text{min}\{\Lambda_1,\Lambda_2,\cdots,\Lambda_N\}$. Therefore,  $\underline{\Lambda}(j\omega)\in\boldsymbol{\Lambda}(j\omega)=[\underline{\Lambda}(j\omega),\cdots,\Lambda_N(j\omega)]\in\mathbb{R}^N$ is the minimum eigenvalue of the hermitian matrix $\bm{\mathcal{H}}(j\omega) = \boldsymbol{G}(j\omega)+\boldsymbol{G}^\dagger(j\omega)$, also referred as passivity index hereafter in this paper. Note that the eigenvalues $\boldsymbol{\Lambda}$ are real due to $\bm{\mathcal{H}}$ being hermitian, and thus condition 2 is equivalent to checking whether $\bm{\mathcal{H}}$ is positive semi-definite \cite{bao2007}.

If two subsystems are connected in a feedback manner and both are passive over a specific frequency range $[\omega_{a},\omega_{b}]$, it implies that there are no instability issues for this frequency region \cite{bao2007}. Since the FD formulation provides a feedback interconnection of the passive network $\boldsymbol{Y^{net}}$ with other devices $\boldsymbol{Y^{a}}$, this property is frequently exploited to passivate the high-frequency region of the converter dynamics, thereby preventing negative interaction with the interconnected system \cite{Chatterjee23,Wang2024,Josep2024}. 
As a result, it allows system operators to focus primarily on lower frequencies for stability analysis. 
Additionally, it is utilized for the control design of the converter to enhance robustness against interactions between the converter and the system by expanding the range of passive frequencies \cite{Dey2021}. 
With this passivity-oriented approach to stability prediction, a system can be expected to remain stable within passivated frequency range where both interconnected systems are passive. 
\subsection{Review of Sensitivity Analysis Methods}
To gain additional insights beyond stability prediction, sensitivity analysis is necessary for any system under study.
The results of this analysis offer valuable guidance for designing stable and more robust systems.
In SS models, the eigenvalue sensitivity with respect to an arbitrary parameter can be described as follows:
\begin{equation}
   \frac{\partial\boldsymbol{\lambda}}{\partial \rho} = \boldsymbol{\psi^A}\frac{\partial \boldsymbol{A}}{\partial\rho}\boldsymbol{\Phi^A}, \quad p_{ki}=\frac{\partial \lambda_i}{\partial a_{kj}}
   \label{eq: SS parametric sensitivity}
\end{equation}
where $\boldsymbol{\psi^A}=[\boldsymbol{\psi^A_1,\cdots,\psi^A_K}]\in\mathbb{C}^{K\times K}$ and $\boldsymbol{\Phi^A}=[\boldsymbol{\Phi^A_1,\cdots,\Phi^A_K}]\in\mathbb{C}^{K\times K}$ denote left and right eigenvectors of the state matrix A with respect to $\boldsymbol{\lambda}$, respectively, the subscript refer to the $i$-th mode, $j$-th column, and $k$-th row and $\rho$ is an arbitrary parameter of the system.
$p_{ki}$ is the state-space participation factor (SS-PF) of $\lambda_i$ to $a_{kj}$, which is an element of $\boldsymbol{A}$. 
From the above equations, it is shown that the PF of the state variable is equivalent to the parametric sensitivity \cite{Kundur}. 
However, utilizing the above SS-PF can be challenging for large-scale systems, especially when constructing a combined full-order model with black-box devices. 
Additionally, this approach cannot be used to prevent instabilities by design and it only provides sensitivity results at the device or parameter level.

To address these issues, FD model-based PF analysis (FD-PF) is employed \cite{Grey-Box}. 
In evaluating FD-PF, the nodal admittance model is diagonalized at specific frequencies or modes of interest as \cite{PMD2021,Orellana2023, Bus-PF, MTHVDCstability}
\begin{equation}    \boldsymbol{\Phi^n}\boldsymbol{\lambda^n}\boldsymbol{\psi^n}=\boldsymbol{Y^n}\big|_{s=j\omega\,\text{or}\,\lambda}
\label{eq: Osc Mode Identification}
\end{equation}
\begin{equation}
    \boldsymbol{P^n}_{ij}=\frac{\partial\lambda_{c}^n}{\partial \boldsymbol{Y^n}_{ij}}=\boldsymbol{\psi^n}\frac{\partial \boldsymbol{Y^n}}{\partial \boldsymbol{Y^n}_{ij}}\boldsymbol{\Phi^n}
    \label{eq: FD-PF}
\end{equation}
where $\boldsymbol{\lambda^n}=\text{diag}(\lambda_1^n,\cdots,\lambda_{2N}^n)\in\mathbb{C}^{2N\times 2N}$ is the diagonal eigenvalues matrix, $\lambda_{c}^n\in\boldsymbol{\Lambda^n}$ is the critical eigenvalue of nodal admittance, $\boldsymbol{Y^n}_{ij}\in\boldsymbol{Y^n}$ is the $i,j$ component of the nodal admittance, $\boldsymbol{P^n}_{ij}$ is the participation factor matrix, with $\boldsymbol{\psi^n}=[\boldsymbol{\psi_1^n,\cdots,\psi_N^n}]\in\mathbb{C}^{2N\times 2N}$ and $\boldsymbol{\Phi^n}=[\boldsymbol{\Phi_1^n,\cdots,\Phi_N^n}]\in\mathbb{C}^{2N\times 2N}$ denoting the left and right eigenvector matrices, respectively. 

This eigenvalue decomposition approach was first used to address harmonic resonance issues \cite{Harmonic-Resonance}.
Based on the singularity of the system, one of the eigenvalues from the system impedance is nearly zero around the resonance frequency. As a result, finding the minimum eigenvalues from \cref{eq: Osc Mode Identification} provides nearly accurate results about the system's oscillation modes.

\begin{figure*}[!t]
\centering
\includegraphics[width=7.0in]{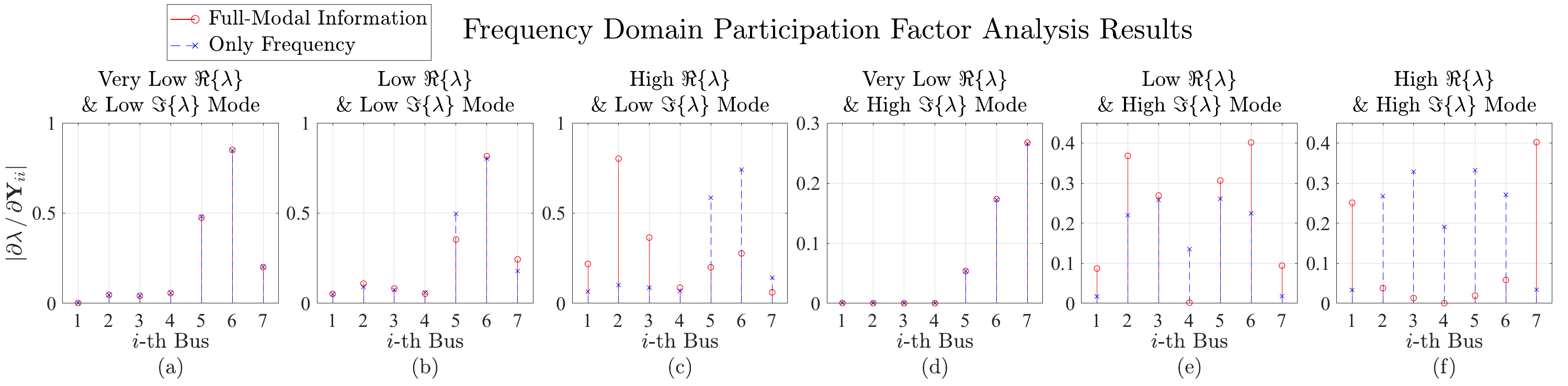}
\caption{Frequency domain sensitivity analysis results for bust elements with full modal information and only frequency cases (a) $\lambda=-0.18\pm j0.29$, $\zeta$ = 1.89 (b) $\lambda=-0.64\pm j8.30$, $\zeta$ = 13.07 (c) $\lambda=-1.98\pm j7.81$, $\zeta$ = 4.06 (d) $\lambda=-40.32\pm j22774.20$, $\zeta$ = 564.84 (e) $\lambda=-31.19\pm j376.84$, $\zeta$ = 12.12 and (f) $\lambda=-1153.58\pm j769.07$, $\zeta$ = 1.2.}
\label{fig: Limitations_of_FD}
\vspace{-0.3cm}
\end{figure*}

However, as previously mentioned, this method may produce inaccurate analysis results when the mode's real part has a significant value or in lower-frequency interaction studies where the required conditions are not easily met, thus losing generality. 
These inaccuracies can be overcome by using exact eigenvalue information rather than relying solely on frequencies response data. 
This limitation is demonstrated in \cref{fig: Limitations_of_FD}, with the eigenvalues corresponding to the case study later introduced in Section IV.
It can  be seen that only cases with a very small real part of the SS mode yield accurate results, indicating that methods utilizing only frequency response data might not accurately capture the sensitivity magnitude compared to using full-modal information.
The results show that it is not directly related to damping ratio. 
Instead, inaccuracy arises due to the non-negligible $\Re\{\lambda\}$, thus leading to a non-zero determinant of $\boldsymbol{Y^n}$. 
This weakens the connection to the critical eigenvalue (a zero eigenvalue of $\boldsymbol{Y^n}$) and leads to inaccurate sensitivity analysis. 
Consequently, the SS-PF and the FD-PF relationship in \cref{eq: FD-PF} does not hold properly.

In contrast, the aforementioned accuracy limitation can be overcome by using full modal information for sensitivity analysis (i.e., exact eigenvalue including $\Re\{\lambda\}$ for substituting $s$).
When exact eigenvalue information is utilized, and sensitivity analysis of critical eigenvalue $\lambda_c^n$ of $\boldsymbol{Y^n}$  shows exact sensitivity magnitudes \cite{On-Relationship}.
It is also called the critical mode sensitivity of $\boldsymbol{Y^n}$ utilizing the fact
\begin{equation}
    \text{det}[\boldsymbol{Y^n}(s)|_{s=\lambda_i}]=0 \quad \forall i
\end{equation}
which implies there must be at least one zero eigenvalue (i.e., $\lambda_c^n$) from $\boldsymbol{Y^n}(\lambda_i)$.

However, although this sensitivity approach accurately reveals the magnitude of sensitivity, it lacks precise directional information \cite{Root-Cause}.
This limitation arises from the missing link between critical mode sensitivity and exact FD eigenvalue sensitivity.
Therefore, due to the absence of a directional insight, critical mode sensitivity may not be sufficient to design a system or controller to improve system performance or stability.

To address these issues, the exact FD eigenvalue sensitivity analysis method is proposed in \cite{Root-Cause, On-Relationship, Oscillatory-Stability}. 
To bridge the gap between critical mode PF and FD eigenvalue sensitivity, a sensitivity analysis method is derived as
\begin{equation}
\frac{\partial \lambda_i}{\partial \boldsymbol{Y_{ij}^n}}=\frac{\partial\lambda_i}{\partial \text{det}[\boldsymbol{Y^n}]}\frac{\partial \text{det}[\boldsymbol{Y^n}]}{\partial \lambda_c^n}\frac{\partial \lambda_c^n}{\partial \boldsymbol{Y_{ij}^n}}=\xi\frac{\partial \lambda^n_c}{\partial \boldsymbol{Y_{ij}^n}} 
\end{equation}
where $\xi$ is a coefficient for compensating the difference between critical mode and FD eigenvalue sensitivity, and it can be evaluated as
\begin{equation}
    \xi=\frac{\partial\lambda_i}{\partial \text{det}[\boldsymbol{Y^n}]}\frac{\partial \text{det}[\boldsymbol{Y^n}]}{\partial \lambda_c^n}=-\frac{\text{tr(adj(}\boldsymbol{Y^n}(\lambda_i)))}{\text{det}[\boldsymbol{Y^n}]'(\lambda_i)}
\end{equation}
where $\text{det}[\boldsymbol{Y^n}]'$ is the derivative of the determinant of nodal admittance with respect to the frequency. $\text{tr}(\cdot)$ and $\text{adj}(\cdot)$ denote the trace and adjugate of a matrix. 
This FD eigenvalue sensitivity method provides complete information enabling a more effective system design.

However, this method assumes that the necessary FD models and eigenvalues are already available. 
In practice, obtaining the required model or full-modal information can be challenging.
Especially, obtaining eigenvalues through the FD model might be difficult for high-order component models and high-dimension nodal admittance matrix, resulting in substantial computational cost associated with finding the roots of the determinant, which might also be prone to numerical inaccuracies.

Therefore, we propose a passivity sensitivity analysis method for stability prediction and interaction mitigation, which can overcome aforementioned challenges.
Compared to currently available sensitivity analysis methods, the proposed passivity sensitivity does not require exact modal information while it can still provide direct insights for system design.
In addition, it is computationally more efficient as it avoids the need to calculate the zeros of the determinant of the matrix.

\section{Proposed Passivity Sensitivity Methodology}
\begin{figure}[!t]
\centering
\includegraphics[width=3.3in]{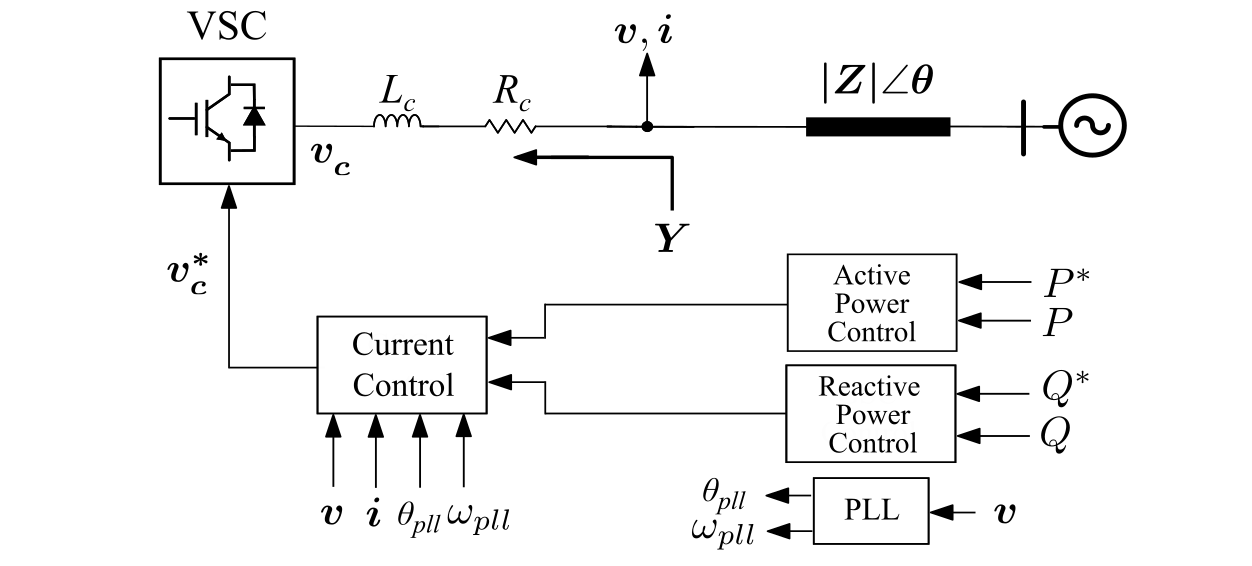}
\caption{GFL VSC scheme.}
\label{fig: GFL_VSC}
\vspace{-0.5cm}
\end{figure}

In this section, the passivity sensitivity of a device with respect to an arbitrary parameter is developed. 
Traditionally, the passivity index has been mainly used for device-level controller design.
However, passivity analysis has been mostly performed by iteratively changing parameters and re-evaluating the index as analytical models are not generally tractable.
In addition, this paper extends previous passivity works to system-level analysis with a passivity sensitivity method and shows its applications with case studies.
The case used for device-level validation focuses on a grid following (GFL) VSC, which is presented in \cref{fig: GFL_VSC}, while a 3-bus system is used for system-level validation, presented in \cref{fig: 3-bus system}.
\subsection{Device-level Passivity Sensitivity}
Parametric passivity sensitivity can be utilized in device control design to relax stability concerns in specific frequency regions (i.e., expanding the dissipative region).
For devices, they can be represented by their frequency responses at each frequency, which contain its terminal dynamic behavior as
\begin{equation}
        \boldsymbol{i}(j\omega)=\boldsymbol{Y}(j\omega)\boldsymbol{v}(j\omega)
\end{equation}
where $\boldsymbol{i}(j\omega)=[i_d(j\omega),i_q(j\omega)]^T\in\mathbb{C}^2$,  $\boldsymbol{v}(j\omega)=[v_d(j\omega),v_q(j\omega)]^T\in\mathbb{C}^2$ which are composed of $d$- and $q$-axis components of voltage and current. 
Device admittance model with each $d$- and $q$ axis is $\boldsymbol{Y}(j\omega) = [Y_{dd}(j\omega), Y_{dq}(j\omega); Y_{qd}(j\omega), Y_{qq}(j\omega)]\in\mathbb{C}^{2\times2}$ which represents the current response $\boldsymbol{i}(j\omega)$ to voltage perturbations $\boldsymbol{v}(j\omega)$. 
Its power can be expressed as \cite{Passivity_input_admittance}
\begin{talign}
    P(j\omega)&=\Re\{v_d(j\omega)i_d(j\omega)^\dagger+v_q(j\omega)i_q(j\omega)^\dagger\}\\
    &=\frac{1}{2}(\boldsymbol{v}^\dagger(j\omega) \boldsymbol{i}(j\omega)+\boldsymbol{i}^\dagger(j\omega)\boldsymbol{v}(j\omega))
    \label{eq: device hermitian}
\end{talign}
where $P(j\omega)$ is power through $\boldsymbol{Y}(j\omega)$. 
superscript $\dagger$ denotes the conjugate-transpose operator.
According to \cref{eq: Passivity Eqns}, it is directly related to system's passivity; therefore, the passivity index of the device's admittance model can be derived as
\begin{talign}
    &\boldsymbol{v}^\dagger(j\omega) \boldsymbol{i}(j\omega)+\boldsymbol{i}^\dagger(j\omega)\boldsymbol{v}(j\omega)\\
    &=\boldsymbol{v}^\dagger(j\omega)\boldsymbol{Y}(j\omega)\boldsymbol{v}(j\omega)+\{\boldsymbol{Y}(j\omega)\boldsymbol{v}(j\omega)\}^\dagger\boldsymbol{v}(j\omega)\\
    &=\boldsymbol{v}^\dagger(j\omega)\{\underbrace{\boldsymbol{Y}(j\omega)+\boldsymbol{Y}^\dagger(j\omega)}_{\bm{\mathcal{H}}(j\omega)}\}\boldsymbol{v}(j\omega)
\end{talign}
\begin{equation}
\bm{\mathcal{H}}(j\omega)=
    \begin{bmatrix}
2\Re\{Y_{dd}(j\omega)\} & Y_{dq}(j\omega) + Y_{qd}^\dagger(j\omega)\\
Y_{qd}(j\omega) + Y_{dq}^\dagger(j\omega) & 2\Re\{Y_{qq}(j\omega)\}
    \end{bmatrix}
    \label{eq: Hermitian of Admittance}
\end{equation}
where $\bm{\mathcal{H}}(j\omega)$ is hermitian matrix by summation of $\boldsymbol{Y}(j\omega)$ and its conjugate transpose. 
If $\bm{\mathcal{H}}(j\omega)$ is positive definite at $\omega$, the device is passive at $\omega$, i.e., $\boldsymbol{\mathcal{H}}(j\omega)$ has a positive minimum eigenvalue or positive passive index, $\underline{\Lambda}\{\bm{\mathcal{H}}(j\omega)\}>0$.
Hereafter, the notation of $(j\omega)$ is skipped if not specified for brevity.
Due to the special structure of $\boldsymbol{\mathcal{H}}$ in \cref{eq: Hermitian of Admittance}, the sensitivity of $\boldsymbol{\mathcal{H}}$ with respect to any $dq$ admittance component $y$ of $\boldsymbol{Y}$ is not straightforward. 
The admittance passivity sensitivity is different depending on whether an off-diagonal or diagonal element is considered. 
To address this complexity, the chain rule is successively applied to develop the passivity sensitivity with respect to an arbitrary parameter $\rho$. 
The resultant parametric passivity sensitivity of the device can be mathematically expressed as \eqref{eq: Parametric Passivity Sensitivity}-\eqref{eq: Case for device pass sens}
\begin{equation}
    \frac{\partial \underline{\Lambda}\{\boldsymbol{\mathcal{H}}\}}{\partial\rho}=\sum_{i,j}\sum_{p,q}\frac{\partial\underline{\Lambda}\{\boldsymbol{\mathcal{H}}\}}{\partial\boldsymbol{\mathcal{H}}_{ij}}\frac{\partial\boldsymbol{\mathcal{H}}_{ij}}{\partial  y_{pq}}\frac{\partial  y_{pq}}{\partial \rho}
    \label{eq: Parametric Passivity Sensitivity}
\end{equation}
\begin{equation}
\frac{\partial\mathcal{H}_{ij}}{\partial y_{pq}}\frac{\partial y_{pq}}{\partial \rho}=
        \begin{cases}
         2\Re\{\frac{\partial y_{pq}}{\partial\rho}\}& \text{if}\,p=q\,\,\&\,\,i=j=p\\
        \frac{\partial y_{pq}}{\partial \rho} & 
        \text{if}\, p\neq q\,\,\&\,\,\{i=p\,\&\,j=q\}\\
        \frac{\partial y_{pq}^\dagger}{\partial \rho} & \text{if}\, p\neq q\,\,\&\,\,\{i\neq p\,\&\,j\neq q\}\\
        0 & \text{otherwise}
    \end{cases}
    \label{eq: Case for device pass sens}
\end{equation}
where the passivity index of $\bm{\mathcal{H}}$ at the frequency of interest $\omega$ is $\underline{\Lambda}\{\boldsymbol{\mathcal{H}}\}\in\boldsymbol{\Lambda}\{\boldsymbol{\mathcal{H}}\}=[\underline{\Lambda},\Lambda_2]\in\mathbb{R}^2$,$y_{pq}\in\boldsymbol{Y}$ with $p\,\&\,q\in[d,q]$ are the $dq$ admittance components, $i\,\&\,j$ are matrix indices, $\partial \underline{\Lambda}\{\boldsymbol{\mathcal{H}}\}/\partial\rho$ represents the parametric passivity sensitivity, and $\partial y_{pq}/\partial \rho$ denotes the sensitivity of the $pq$ element of the $dq$ admittance with respect to the parameter $\rho$. 
Expression \cref{eq: Case for device pass sens} reflects the special structure of $\boldsymbol{\mathcal{H}}$  \cref{eq: Hermitian of Admittance} which is exploited in the derivation of the parametric passivity sensitivity. 
Depending on the perturbed $dq$ admittance component, the impact on the sensitivity is different. 
If $y_{pq}$ is in the diagonal then $p=q$, otherwise for off-diagonal elements $p\neq q$, which depends on the index $i,j$ with element of $\boldsymbol{\mathcal{H}}$.
\begin{figure}[!t]
\centering
\includegraphics[width=3.4in]{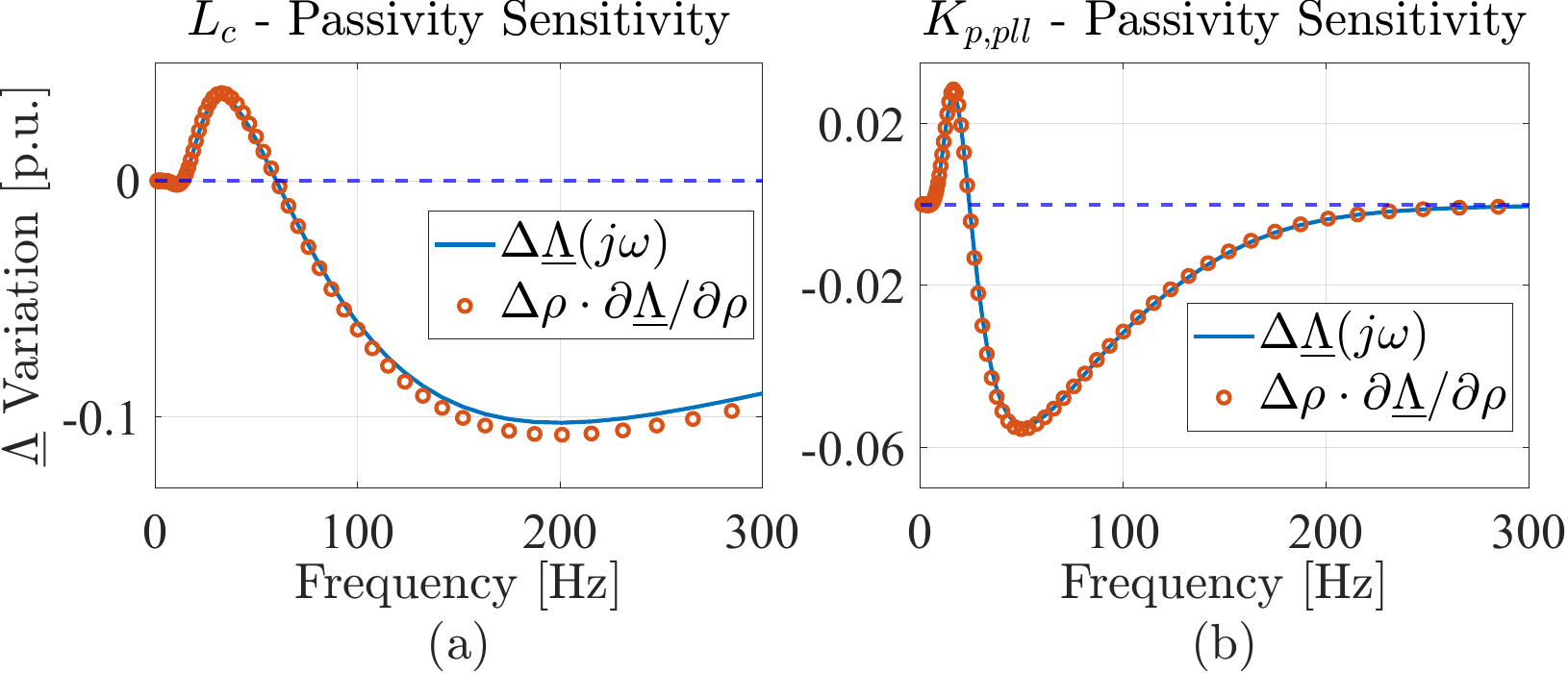}
\caption{Device parametric passivity sensitivity validation results (a) Converter interface inductance $L_{c}$ case, (b) Converter PLL proportional gain $K_{p,pll}$ case}
\label{fig: Device Pass sens val}
\vspace{-0.3cm}
\end{figure}
\begin{figure}[!t]
\centering
\includegraphics[width=3.4in]{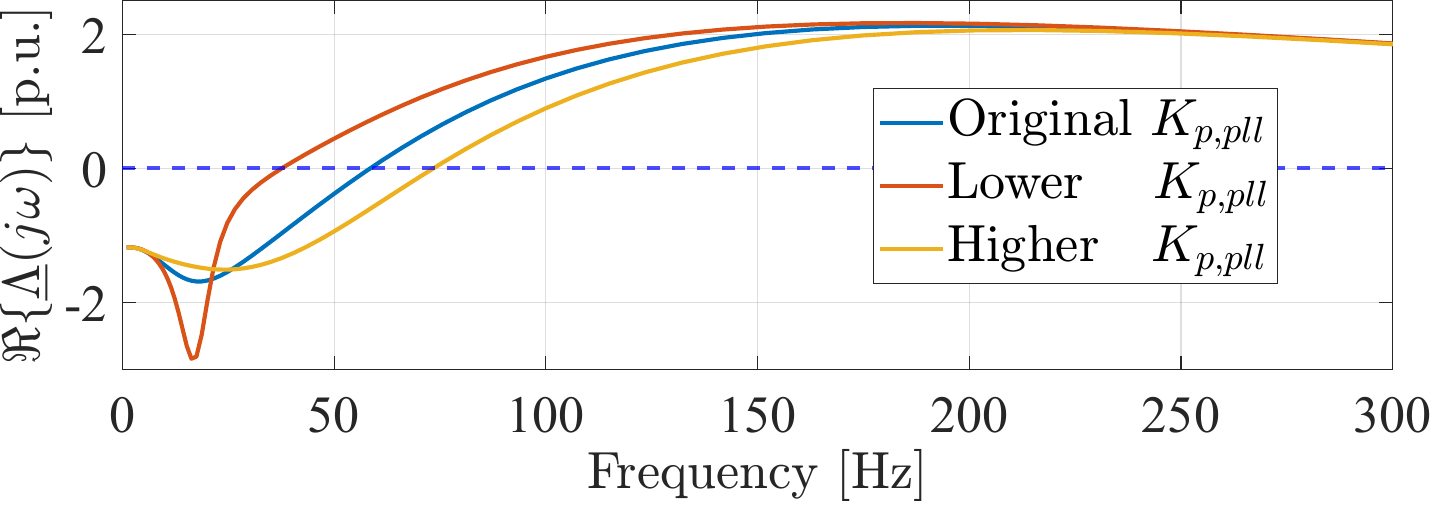}
\caption{Impact of $K_{p,pll}$ on the GFL-VSC's passivity index.}
\label{fig: PLL impacts on Passivity}
\vspace{-0.3cm}
\end{figure}

The remaining part for parametric passivity sensitivity with respect to $\mathcal{H}_{ij}$ can be calculated as
\begin{equation}
    \frac{\partial\boldsymbol{\Lambda}\{\boldsymbol{\mathcal{H}}\}}{\partial \mathcal{H}_{ij}}=\boldsymbol{\psi}\frac{\partial\boldsymbol{\mathcal{H}}}{\partial\mathcal{H}_{ij}}\boldsymbol{\Phi}
    \label{eq: eigen sens Hermitian}
\end{equation}
where $\boldsymbol{\psi}$ and $\boldsymbol{\Phi}$ are left and right eigenvector matrices. This is a similar approach to \cref{eq: SS parametric sensitivity}. 
As a result, the generic parametric passivity sensitivity of the device can be calculated by combining \cref{eq: Parametric Passivity Sensitivity} with the minimum eigenvalue sensitivity analysis results extracted from \cref{eq: eigen sens Hermitian}. 
It enables an effective design of control or device through informative sensitivity analysis results, which reveal not only the magnitude of the parameter's influence but also its directional impact on the overall passivity of the device.
This sensitivity calculation is validated for the converter in \cref{fig: GFL_VSC} in terms of the converter filter inductance $L_c$ and the proportional gain of the phase-locked loop (PLL) $K_{p,pll}$ with respect to a 5\% numerical perturbation of said parameters presented in \cref{fig: Device Pass sens val}.
From \cref{fig: Device Pass sens val}, it can be seen that the numerical perturbation (blue solid lines) and the first-order approximation in terms of the developed sensitivity (orange circles) successfully match. Furthermore, increasing $L_c$ positively influences the passivity index in a broader low-frequency region compared to $K_{p,pll}$. This means that to improve passivity up to around 50 Hz, increasing $L_c$ is more effective than $K_{p,pll}$. In addition, it can be observed that increasing $K_{p,pll}$ negatively affects passivity between 20~Hz and 200~Hz while it has a positive effect up to around 20~Hz.
\cref{fig: PLL impacts on Passivity} reconfirms the sensitivity results with three values of $K_{p,pll}$, i.e. 0.4 p.u. corresponding to the original tuning in blue, a lower value of 0.14 p.u. and a higher value of 0.66 p.u.
\begin{figure}[!t]
\centering
\includegraphics[width=3.4in]{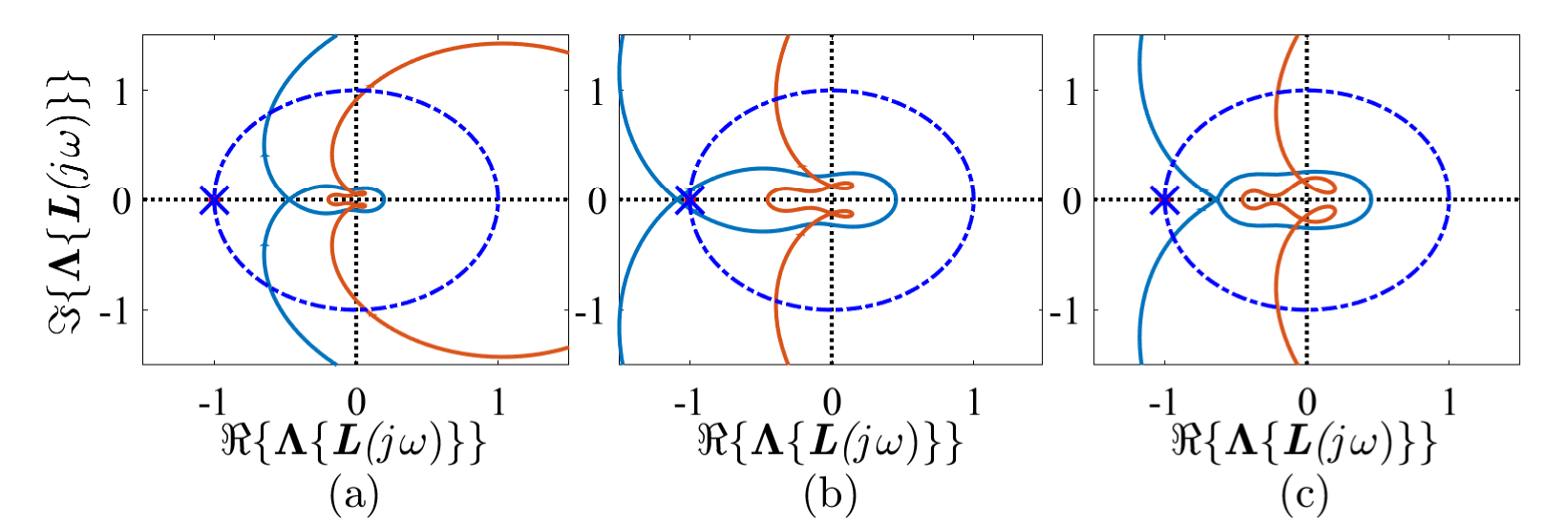}
\caption{Nyquist plots of $\boldsymbol{\Lambda}\{\boldsymbol{L}(j\omega)\}$: (a) $\text{SCR}=3$, base $K_{p,pll}$, (b) $\text{SCR}=1.3$, higher $K_{p,pll}$, (c) $\text{SCR}=1.3$, lower $K_{p,pll}$.}
\label{fig: GNC GFL}
\vspace{-0.3cm}
\end{figure}

\begin{figure}[!t]
\centering
\includegraphics[width=3.3in]{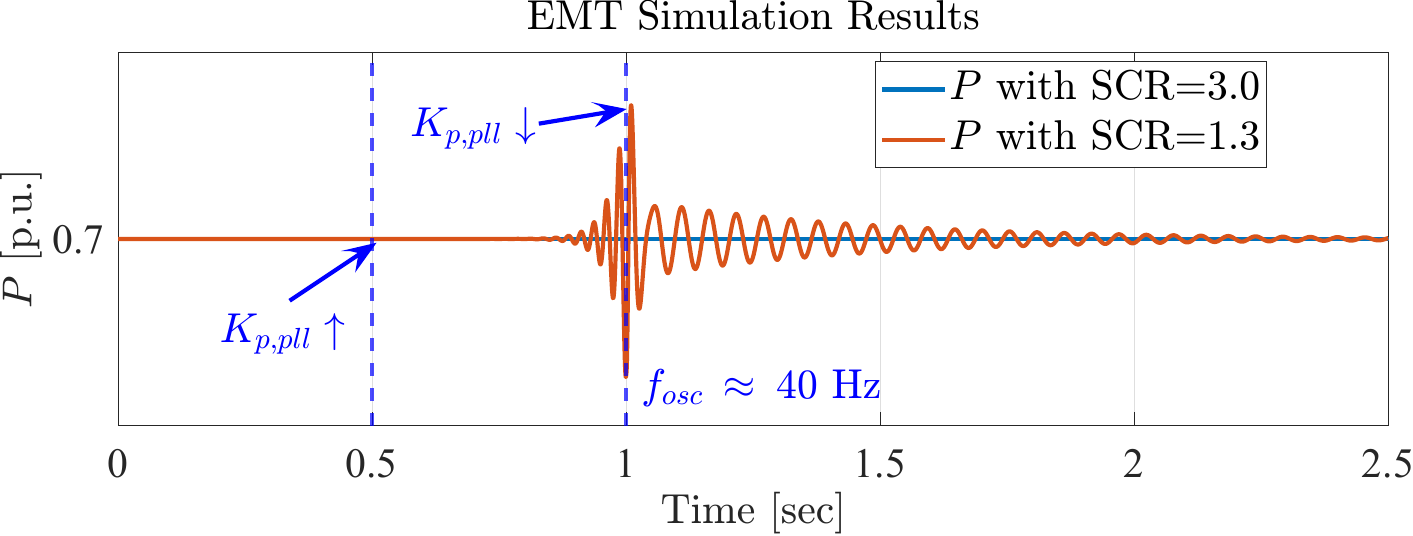}
\caption{EMT simulation results for GFL VSC with various $K_{p,pll}$ and SCR values.}
\label{fig: EMT GFL}
\vspace{-0.3cm}
\end{figure}

To further demonstrate the application to VSC passivation and system stabilization, a widely studied instability resulting from the interaction between a weak grid and a high PLL bandwidth is considered using the system in \cref{fig: GFL_VSC}.
The system is stable with $\text{SCR}=3$ and $\text{X}/\text{R}\approx 6$ as seen by the GNC in \cref{fig: GNC GFL} (a). Next, the grid impedance is increased so $\text{SCR} = 1.3$ and the two previous PLL proportional gain $K_{p,pll}$ values are tested: higher and lower than the original. The GNC in \cref{fig: GNC GFL} (b) and (c) indicates that the former is unstable while the latter is stable.
These stability analysis results are confirmed by EMT simulation for the three PLL gains and two different SCR in \cref{fig: EMT GFL}. As expected, the observed 40 Hz oscillatory instability falls within the non-passive region of the VSC.
It is worth noting that the passivity index value in the non-passive region does not provide further stability insights; i.e., non-passivity does not imply instability.
\subsection{System-level Passivity Sensitivity}
In this subsection, system-level passivity sensitivity is developed, which differs from device-level parametric passivity sensitivity. 
Although parametric passivity sensitivity is useful for analyzing individual devices, passivating all devices in the system might be challenging and overly conservative as the passivity contribution from the network and interaction between components are overlooked.
In contrast, the proposed system-level passivity sensitivity analysis offers a less conservative approach while achieving the required closed-loop passivity by exploiting system-level information.

For expanding the device-level parametric sensitivity analysis to a system-level sensitivity, its characteristic features of the nodal admittance formulation need to be considered, e.g., off-diagonal element of $Y^{n}_{ij}$ will be reflected on both $\mathcal{H}^n_{ij}$, $\mathcal{H}^n_{ji}$, $\mathcal{H}^n_{ii}$, and $\mathcal{H}^n_{jj}$.
The nodal admittance can be expressed as
\begin{equation}
    \boldsymbol{Y^n}_{ij}=
\begin{cases}
    \boldsymbol{Y}_{ij} + \sum_{i\neq j} \boldsymbol{Y}_{ij} & \text{if}\,\, i=j\\
    -\boldsymbol{Y}_{ij} & \text{if}\,\, i\neq j
\end{cases}
\label{eq: Ynodal}
\end{equation}
where $\boldsymbol{Y}_{ij}$ is admittance of device or branch between bus $i$ and $j$.
For properly deriving the system-level passivity sensitivity, the system nodal hermitian matrix $\boldsymbol{\mathcal{H}^n}$ is constructed using the nodal admittance $\boldsymbol{Y^n}$ in a similar fashion as \cref{eq: device hermitian},\\
\begin{talign}
    \boldsymbol{\mathcal{H}^n} &= \boldsymbol{Y^n}+\boldsymbol{Y^{n\dagger}}\\ 
    &=
    \begin{bmatrix}
        \boldsymbol{Y_{11}^n}+(\boldsymbol{Y_{11}^{n\dagger}}) & \cdots & \boldsymbol{Y_{1N}^n}+(\boldsymbol{Y_{N1}^{n^\dagger}}) \\
        \vdots & \ddots & \vdots \\
        \boldsymbol{Y_{N1}^n}+(\boldsymbol{Y_{1N}^{n^\dagger}}) & \cdots & \boldsymbol{Y_{NN}^n}+(\boldsymbol{Y_{NN}^{n^\dagger}})
    \end{bmatrix}
    \label{eq: Hermitian of nodal}
\end{talign}
\begin{equation}
    \boldsymbol{Y^n}=
    \begin{bmatrix}
        \boldsymbol{Y_{11}^n} & \cdots & \boldsymbol{Y_{1N}^n} \\
        \vdots & \ddots & \vdots \\
        \boldsymbol{Y_{N1}^n} & \cdots & \boldsymbol{Y_{NN}^n}
    \end{bmatrix}
\end{equation}
\begin{equation}
    \boldsymbol{Y^{n^\dagger}}=
    \begin{bmatrix}
        (\boldsymbol{Y_{11}^{n\dagger}}) & \cdots & (\boldsymbol{Y_{N1}^{n\dagger}}) \\
        \vdots & \ddots & \vdots \\
        (\boldsymbol{Y_{1N}^{n\dagger}}) & \cdots & (\boldsymbol{Y_{NN}^{n\dagger}})
    \end{bmatrix}
\end{equation}
where $\boldsymbol{Y_{ij}^n}$ is the $i$-th row and $j$-th column component of $\boldsymbol{Y^n}$ and $\boldsymbol{\mathcal{H}^n}$ is the hermitian matrix of $\boldsymbol{Y^n}$.
From \cref{eq: Ynodal} and \cref{eq: Hermitian of nodal}, it can be observed that if the off-diagonal element of $\boldsymbol{Y^n}$ is changed, it is reflected on not only its off-diagonal but also diagonal component. 
Due to this feature, the perturbation of elements in $\boldsymbol{Y^n}$ from $\boldsymbol{Y^{sh}}$ and $\boldsymbol{Y^e}$ need different cautions. 

Therefore, the variation in $\Delta\boldsymbol{Y^n}$ with respect to $\Delta\boldsymbol{Y}$ when this perturbation is from the $k$-th branch $\Delta\boldsymbol{Y_k^b}$ can be classified depending on the perturbing element as
\begin{align}
    \Delta\boldsymbol{Y^n}&=
    \begin{cases}
        \boldsymbol{E_{ij}^n}\,\otimes\,\Delta\boldsymbol{Y}& \text{if}\,\,\exists\Delta\boldsymbol{Y^{sh}}\\
        \boldsymbol{\mathcal{I}_i}^T(\boldsymbol{E_k^b}\,\otimes\,\Delta\boldsymbol{Y^b_k})\boldsymbol{\mathcal{I}_i}& \text{if}\,\,\exists\Delta\boldsymbol{Y^e}
    \end{cases}
    \label{eq: nodal variation}
\end{align}
where $\otimes$ denotes Kronecker product, $\boldsymbol{E_{ij}^n}\in\mathbb{R^{N\times N}}$ is a matrix unit, and the system network topology and where the perturbation comes from determine it. 
$\boldsymbol{Y}$ can be any device admittance model of system such as $\boldsymbol{Y^{a,b,c}}$. 
$\boldsymbol{E_{k}^b}\in\mathbb{R}^{B\times B}$ denotes matrix unit which has 1 at $k$-th diagonal component. $\boldsymbol{Y_k^b}$ denotes $k$-th branch admittance of $\boldsymbol{Y^b}$. 
In the case of variation existing at $i\neq j$, the variation is induced by $\boldsymbol{Y^b}$. 
If variation existing at $i=j$, it is induced by one of $\boldsymbol{Y^c}$, $\boldsymbol{Y^a}$, and $\boldsymbol{Y^b}$. To reflect this characteristic feature from $\boldsymbol{Y^n}$ on $\boldsymbol{\mathcal{H}^n}$ so to develop the passivity sensitivity, further calculations are required. 
If there is a small variation in $\Delta \boldsymbol{Y_{ij}^n}$, it leads to variations in $\Delta\boldsymbol{Y^n}$, $\Delta\boldsymbol{Y^{n\dagger}}$, and $\Delta\boldsymbol{\mathcal{H}^n}$. 
However, this $\Delta \boldsymbol{Y^n_{ij}}$ also depends on its perturbed component.
The resultant passivity sensitivity with consideration of $\boldsymbol{\mathcal{H}^n}$ with respect to an arbitrary admittance perturbation can be expressed as
\begin{equation}
    \frac{\partial \underline{\Lambda}^n}{\partial \boldsymbol{\mathcal{H}^n}}
    \frac{\partial \boldsymbol{\mathcal{H}^n}}{\partial \boldsymbol{Y^n}}
    \frac{\partial \boldsymbol{Y^n}}{\partial \boldsymbol{Y^n_{ij}}}
    \frac{\partial \boldsymbol{Y^n_{ij}}}{\partial \boldsymbol{Y}}
    =
    \frac{\partial \underline{\Lambda}^n}{\partial \boldsymbol{Y}}
\end{equation}
where the nodal passivity index is $\underline{\Lambda}^n\in\boldsymbol{\Lambda}^n=[\underline{\Lambda}^n,\cdots,\Lambda^n_N]\in\mathbb{R}^N$. In addition, $\Delta \boldsymbol{Y^n}$ due to $\Delta \boldsymbol{Y}$ can be expressed as
\begin{equation}
    \Delta \boldsymbol{Y^n}
    =
    \sum^N_{ij}
    \frac{\partial \boldsymbol{Y^n}}{\partial \boldsymbol{Y^n_{ij}}}
    \frac{\partial \boldsymbol{Y^n_{ij}}}{\partial \boldsymbol{Y}}
    \Delta\boldsymbol{Y}
\end{equation}
\begin{figure}[!t]
\centering
\includegraphics[width=2.8in]{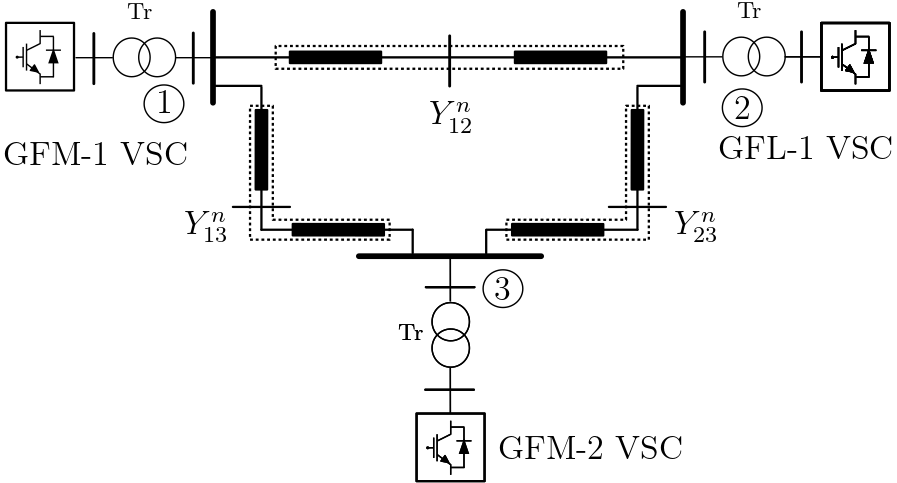}
\caption{3 Bus system for nodal passivity sensitivity validation.}
\label{fig: 3-bus system}
\vspace{-0.3cm}
\end{figure}
\begin{figure}[!t]
\centering
\includegraphics[width=3.4in]{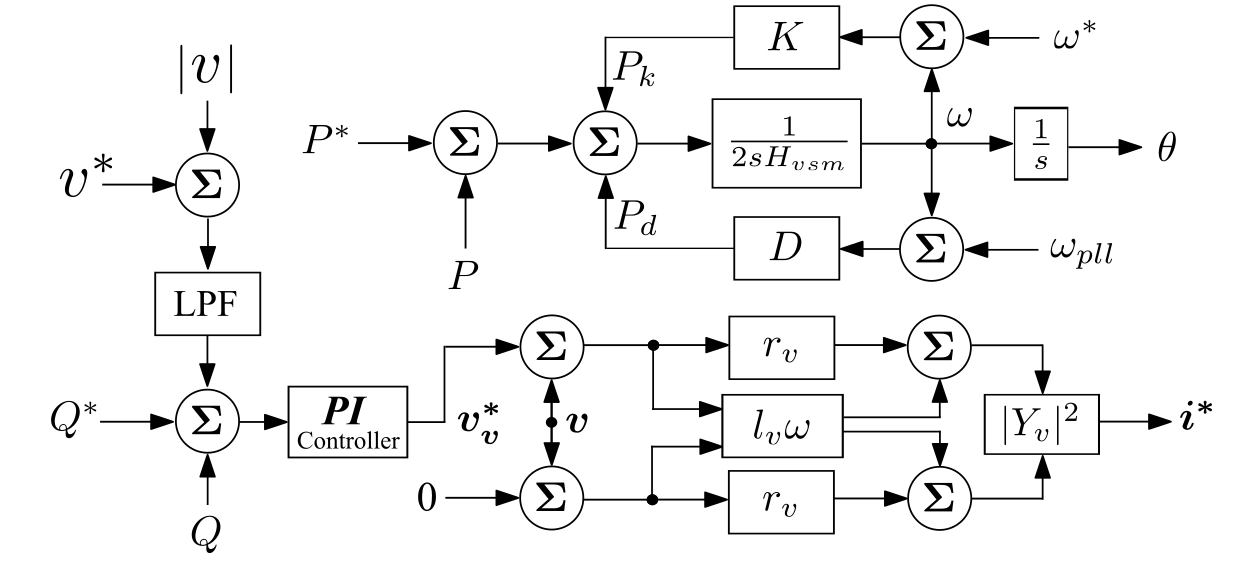}
\caption{GFM VSC control scheme.}
\label{fig: GFM Control}
\vspace{-0.3cm}
\end{figure}
This nodal admittance variation leads to $\Delta \boldsymbol{\mathcal{H}^n}$ as
\begin{talign}
\boldsymbol{\mathcal{H}^n}+\Delta\boldsymbol{\mathcal{H}^n}&=(\boldsymbol{Y^n}+\Delta\boldsymbol{Y^n})+(\boldsymbol{Y^n}+\Delta\boldsymbol{Y^n})^\dagger
    \\&=(\boldsymbol{Y^n}+\boldsymbol{Y^{n\dagger}})+(\Delta\boldsymbol{Y^n}+\Delta\boldsymbol{Y^{n\dagger}})
\end{talign}
\begin{equation}\Delta\boldsymbol{\mathcal{H}^n}=\Delta\boldsymbol{Y^n}+\Delta\boldsymbol{Y^{n\dagger}}
\label{eq: nodal hermitian variation}
\end{equation}
 $\partial \boldsymbol{Y^n}/\partial\boldsymbol{Y}$ can be found from \cref{eq: nodal variation}.
$\Delta \boldsymbol{\mathcal{H}^n}$ can be calculated using \cref{eq: nodal variation} and \cref{eq: nodal hermitian variation} for each perturbed component case as
\begin{talign}
    \Delta\boldsymbol{\mathcal{H}^n}&=\Delta\boldsymbol{Y^n} + \Delta\boldsymbol{Y^{n\dagger}}\quad\text{if}\,\,\exists\Delta\boldsymbol{Y^{sh}}\\
    &=\boldsymbol{E_{ij}^n}\,\otimes\, \Delta\boldsymbol{Y} + (\boldsymbol{E_{ij}^n}\,\otimes\, \Delta\boldsymbol{Y})^\dagger\quad\quad\quad\quad\quad\\
    &=\boldsymbol{E_{ij}^n}\,\otimes\, (\Delta\boldsymbol{Y} + \Delta\boldsymbol{Y}^\dagger)
\end{talign}
\begin{talign}
    \Delta\boldsymbol{\mathcal{H}^n}&=\Delta\boldsymbol{Y^n} + \Delta\boldsymbol{Y^{n\dagger}}\quad\text{if}\,\,\exists\Delta\boldsymbol{Y^e}\\
    &=\boldsymbol{\mathcal{I}_i}^T(\boldsymbol{E_k^b}\,\otimes\,\Delta\boldsymbol{Y^b_k})\boldsymbol{\mathcal{I}_i}
    +
    (\boldsymbol{\mathcal{I}_i}^T(\boldsymbol{E_k^b}\,\otimes\,\Delta\boldsymbol{Y^b_k})\boldsymbol{\mathcal{I}_i})^\dagger\\
    &=\boldsymbol{\mathcal{I}_i}^T((\boldsymbol{E_k^b}\,\otimes\,\Delta\boldsymbol{Y^b_k})
    +(\boldsymbol{E_k^b}\,\otimes\,\Delta\boldsymbol{Y^b_k})^\dagger) \boldsymbol{\mathcal{I}_i}\\
    &=\boldsymbol{\mathcal{I}_i}^T(\boldsymbol{E_k^b}\,\otimes\,(\Delta\boldsymbol{Y^b_k} + \Delta\boldsymbol{Y^{b\dagger}_k}))\boldsymbol{\mathcal{I}_i}\\ 
\end{talign}
where $\boldsymbol{\mathcal{I}_i}^\dagger=\boldsymbol{\mathcal{I}_i}^T$, $\boldsymbol{E^{b\dagger}_k}=\boldsymbol{E^{b T}_k}=\boldsymbol{E^b_k}$, and $\boldsymbol{E^{n\dagger}_{ij}}=\boldsymbol{E^{n T}_{ij}}=\boldsymbol{E^n_{ij}}$ as real-valued matrices. Therefore, $\Delta\boldsymbol{\mathcal{H}^n}$ also can be classified with respect to the perturbed components as
\begin{equation}
    \Delta\boldsymbol{\mathcal{H}^n}=
    \begin{cases}
        \boldsymbol{E_{ij}^n}\,\otimes\, (\Delta\boldsymbol{Y} + \Delta\boldsymbol{Y}^\dagger)& \text{if}\,\,\exists\Delta\boldsymbol{Y^{sh}}\\
        \boldsymbol{\mathcal{I}_i}^T(\boldsymbol{E_k^b}\,\otimes\,(\Delta\boldsymbol{Y^b_k} + \Delta\boldsymbol{Y^{b\dagger}_k}))\boldsymbol{\mathcal{I}_i}& 
        \text{if}\,\,\exists\Delta\boldsymbol{Y^e}
    \end{cases}
    \label{eq: cases nodal hermitian variation}
\end{equation}
For finding the resultant nodal passivity variation $\Delta \boldsymbol{\underline{\Lambda}^n}$, $\Delta \boldsymbol{\Lambda^n}$ can be expressed using $\boldsymbol{\mathcal{H}^n}$'s eigenvalue sensitivity via decomposition similar to \cref{eq: eigen sens Hermitian} as
\begin{equation}
    \Delta \boldsymbol{\Lambda^n}=\sum^N_{ij}
    \boldsymbol{\psi^n}\frac{\partial\boldsymbol{\mathcal{H}^n}}{\partial\mathcal{H}_{ij}^n}\boldsymbol{\Phi^n}
    \Delta\boldsymbol{\mathcal{H}_{ij}^n}
    \label{eq: nodal hermitian sensitivity}
\end{equation}
where $\Delta\boldsymbol{\mathcal{H}^n_{ij}}$ can be found from \cref{eq: cases nodal hermitian variation}. As a result, $\Delta \boldsymbol{\underline{\Lambda}^n}$ can be computed from \cref{eq: nodal hermitian sensitivity}. 

Using the developed calculations described above, the system-level passivity sensitivity can be evaluated for any system with respect to any component or aggregation thereof. 
Compared to the FD eigenvalue sensitivity, this systematic passivity sensitivity gives the required information for identifying the most influential components and improving system stability without necessitating an exact system eigenvalue representation. 
If the system is too large to calculate its eigenvalues from admittance models or black-box models are involved, this method is more attractive due to its simple frequency-wise eigenvalue calculation as opposed to the finding how the roots of the determinant of $\boldsymbol{Y^n}$ change under different measures.

\begin{figure}[!t]
\centering
\includegraphics[width=3.4in]{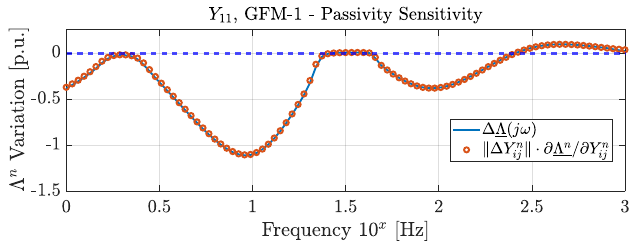}
\caption{Nodal passivity sensitivity validation with GFM-1.}
\label{fig: 3-bus validation}
\vspace{-0.3cm}
\end{figure}

The developed nodal passivity sensitivity is validated for the system in \cref{fig: 3-bus system}, where the GFM VSC's control scheme is shown in \cref{fig: GFM Control}.
The validation involving GFM-1 at bus 1 is shown in \cref{fig: 3-bus validation}. 
\begin{figure}[!t]
\centering
\includegraphics[width=3.4in]{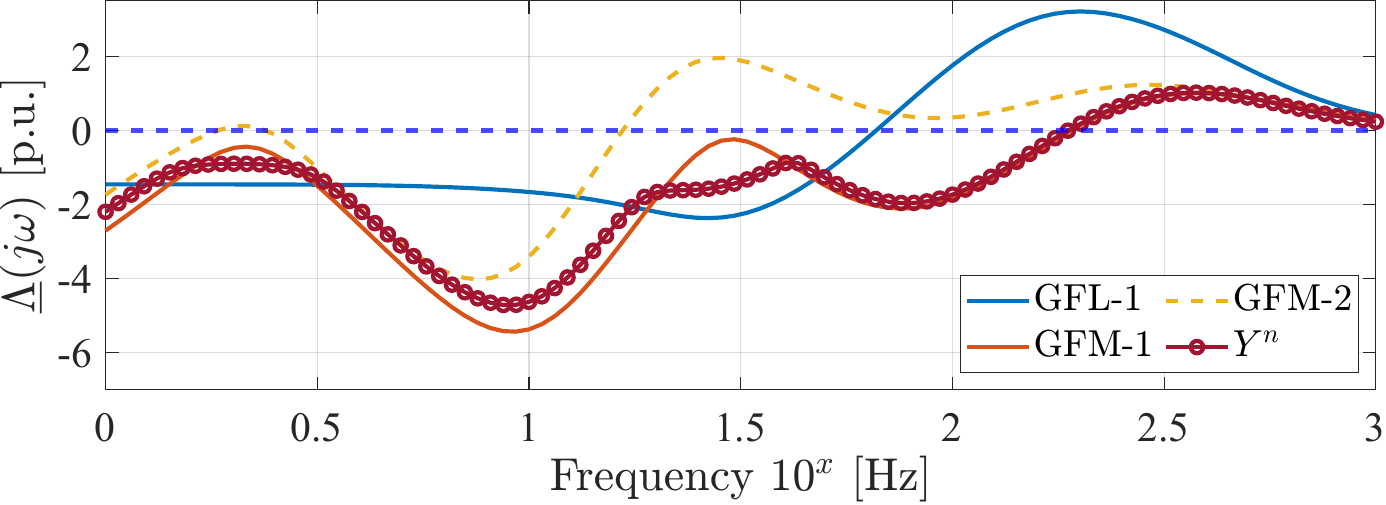}
\caption{Passivity investigation of 3 bus system before passivation.}
\label{fig: 3-bus NP Case}
\vspace{-0.3cm}
\end{figure}
The result shows that the developed sensitivity is quite accurate with a perturbation size $\Delta\boldsymbol{Y}_{ij}$ of 10\% the admittance magnitude.
In addition, it shows that $\underline{\Lambda}^n$ can be improved between [$10^{2.5}$Hz, $10^3$Hz] when the admittance of GFM-1 is proportionally increased and relatively degraded in other frequencies.

\begin{figure}[!t]
\centering
\includegraphics[width=3.4in]{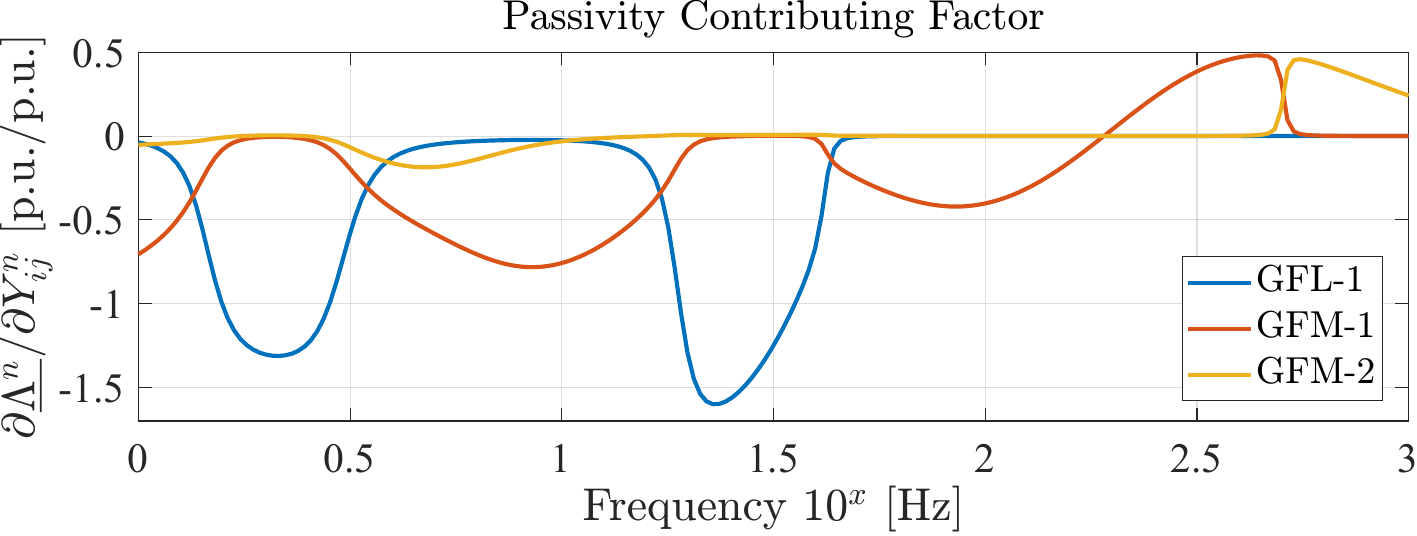}
\caption{System level passivity participation factor.}
\label{fig: 3 bus Passivity contribution}
\vspace{-0.3cm}
\end{figure}

The system's passivity indices are investigated in \cref{fig: 3-bus NP Case}.
The results show that $\underline{\Lambda}^n$ is mainly dominated by GFL-1 and GFM-1 as GFM-2 is relatively more passivated than others.
In addition, the dominant influence on $\underline{\Lambda}^n$ also depends on frequency as they show different passivity over the frequencies.
To further investigate how each component's admittance can contribute to $\underline{\Lambda}^n$ at different frequency ranges, the system level passivity sensitivity $\partial\underline{\Lambda}^n/\partial\boldsymbol{Y_{ij}^n}$ is calculated and presented in \cref{fig: 3 bus Passivity contribution}. 
\begin{figure}[!t]
\centering
\includegraphics[width=3.4in]{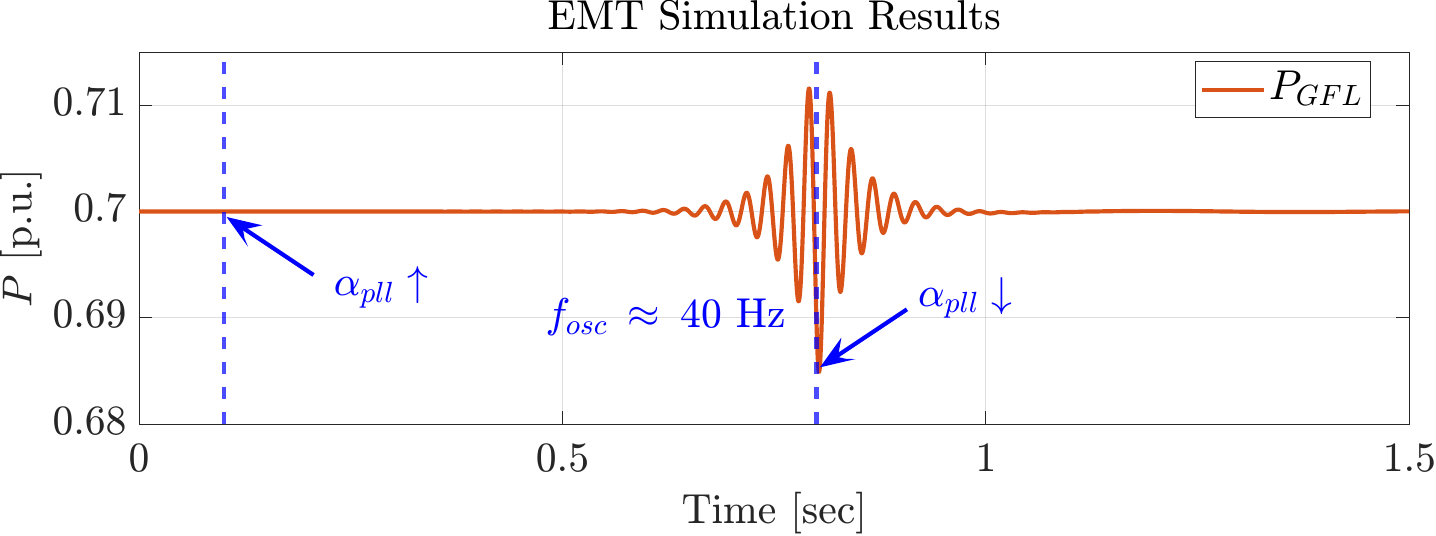}
\caption{EMT simulation results with 3 bus system.}
\label{fig: 3 bus EMT}
\vspace{-0.3cm}
\end{figure}
The results show not only the intensity of contribution but also provide directional information on whether a given component influences $\underline{\Lambda}^n$ positively or negatively.  
In addition, it does not show mode-specific or frequency-oriented results but rather the influence over entire frequency regions.
It reconfirms the results of \cref{fig: 3-bus NP Case} as both show that GFL-1 negatively influences around 45~Hz$\,\,\approx10^{1.6}$~Hz.
However, the result of GFM-1 shows that even though it is not passivated around that frequency band, it does not contribute significantly to the system's passivity index.
These insights together with \cref{fig: 3-bus NP Case} imply that the stability risks in this frequency range are dominated by GFL-1.

\begin{figure}[!t]
\centering
\includegraphics[width=3.3in]{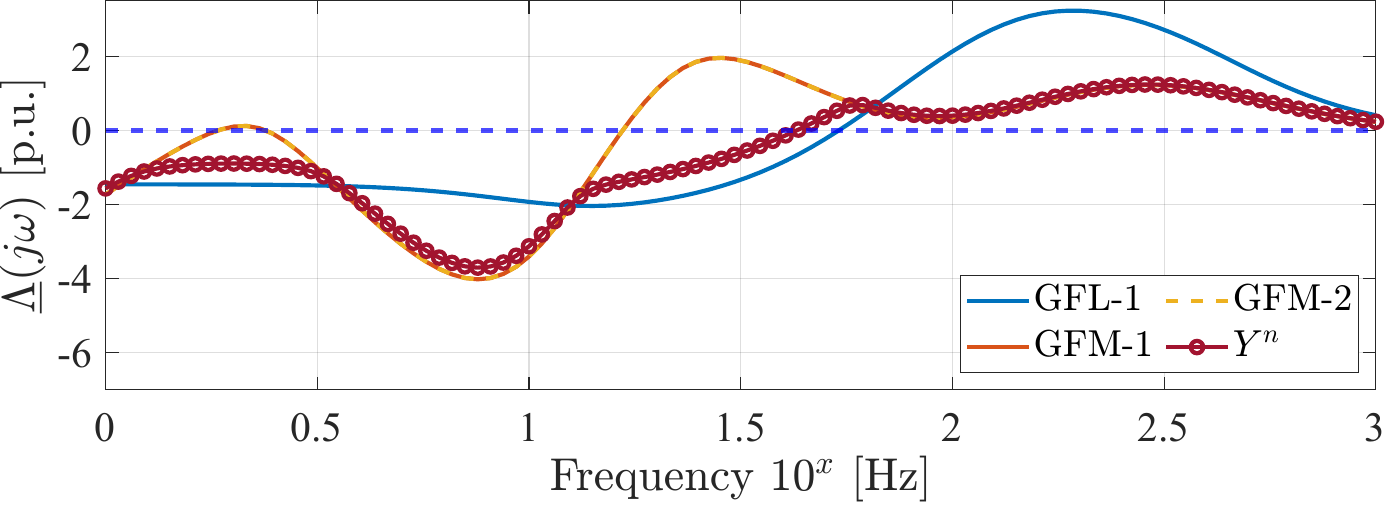}
\caption{Passivity investigation with 3 bus system after passivation.}
\label{fig: 3 bus P}
\vspace{-0.3cm}
\end{figure}

\begin{figure}[!t]
\centering
\includegraphics[width=3.5in]{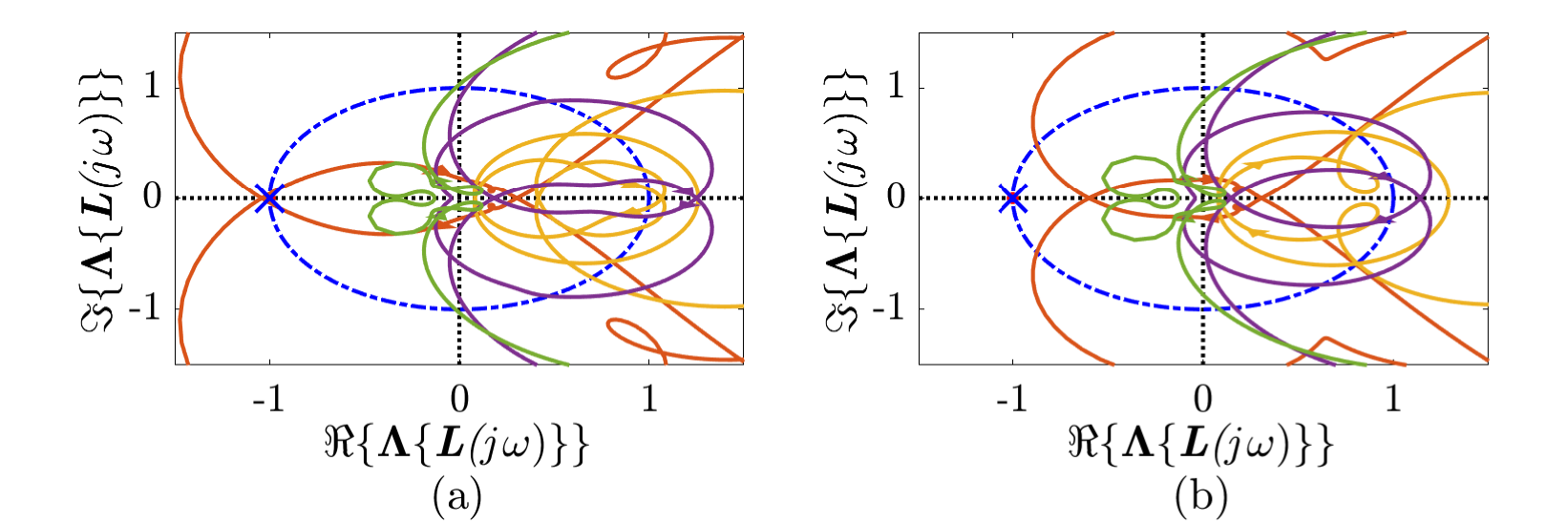}
\caption{Nyquist plots of $\boldsymbol{\Lambda}\{\boldsymbol{L}(j\omega)\}$ with (a) before passivation, (b) after passivation.}
\label{fig: 3 bus GNC}
\vspace{-0.3cm}
\end{figure}

To reconfirm the observation from \cref{fig: 3 bus Passivity contribution}, EMT simulation is performed where GFM-1 is also passivated as GFM-2 and the result is shown in \cref{fig: 3 bus EMT}.
It can be observed that there is still an instability around $f_{osc}\approx40$~Hz with a higher bandwidth of PLL, which is used in \cref{fig: 3-bus NP Case}, even though GFM-1 is also passivated in that frequency band.
However, the system can be stabilized after the bandwidth of PLL is decreased.
The passivity investigation results with decreased bandwidth of PLL and passivated GFM-1 are shown in \cref{fig: 3 bus P}. 
These results show that even though GFL-1 is not fully passivated in the frequency band around 40~Hz, the whole system is effectively passivated in said range as seen by the red curve $\underline{\Lambda}^n$ in \cref{fig: 3 bus P}.
Based on these results, it can be concluded that system-level passivation can be achieved in a less conservative manner by passivity sensitivity analysis. This is especially relevant, as many devices can neither be fully passivated nor need to be so as to achieve system-level passivity and stability.
Therefore, the proposed approach is useful to achieve the design of a more stable system with less effort by investigating $\underline{\Lambda}^n$ and its sensitivities.
These stabilizing effects brought by the passivity approach are also investigated through GNC in \cref{fig: 3 bus GNC}.
These results are aligned with \cref{fig: 3 bus EMT}.
While comparing the single GFL connected to a weak grid case in \cref{fig: GNC GFL} with the 3-bus systems GNC in \cref{fig: 3 bus GNC}, it can be noticed that the number of eigenvalues increases and thus obtaining stability conclusions from the GNC analysis might get more complex as the dimensions of the loop gain increase with the system size.
In contrast, the passivated frequency region ensures stability in said frequency range without the need for a comprehensive investigation of Nyquist plots.

\section{Case Study}
\begin{figure}[!t]
\centering
\includegraphics[width=3.5in]{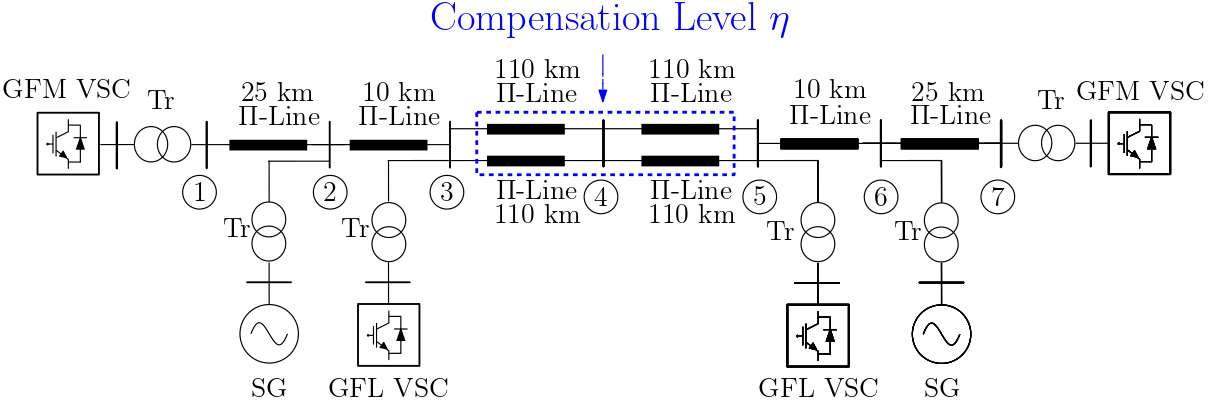}
\caption{Modified two-area system with series compensation.}
\label{fig: 2 area system}
\vspace{-0.3cm}
\end{figure}

\begin{figure}[!t]
\centering
\includegraphics[width=3.5in]{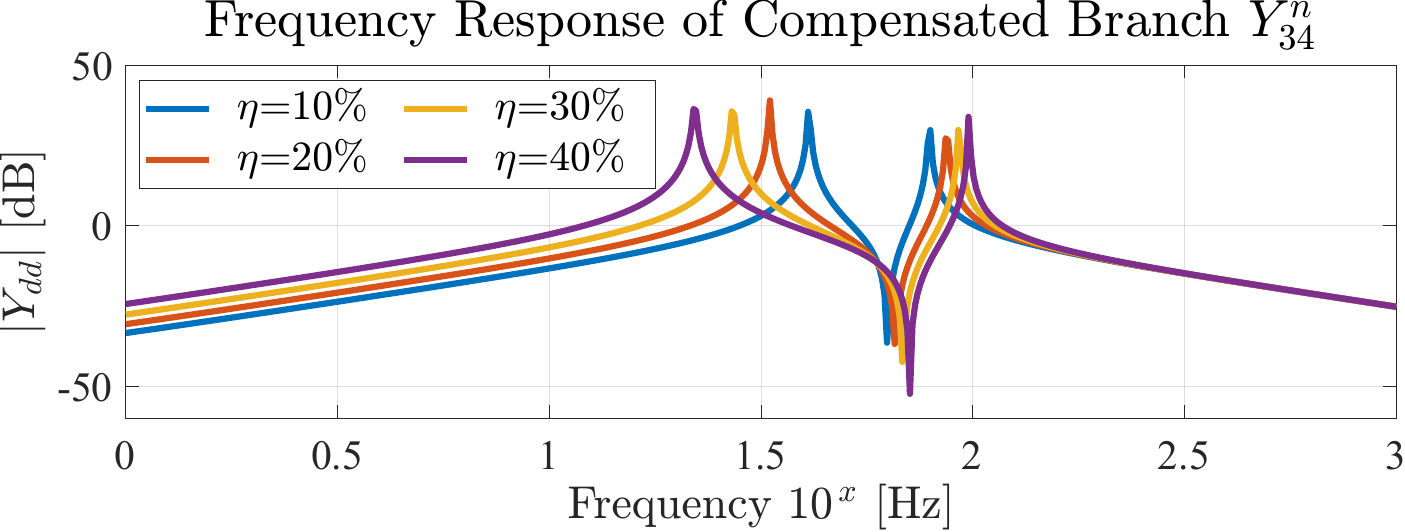}
\caption{Frequency response of compensated network branch.}
\label{fig: Freq response compensated branchh}
\vspace{-0.5cm}
\end{figure}

To demonstrate the applicability of the passivity sensitivity methodology to more complex power systems, a modified two-area system is constructed based on \cite{Kundur}.
The single-line diagram is presented in \cref{fig: 2 area system}, where the loads have been replaced by GFL-VSC and two SGs are replaced by GFM-VSC. In addition, the inter-area lines are compensated by means of series capacitors.
The series compensation strategy is often utilized to improve the power transfer capability \cite{Series_Compensation}. 
However, this compensation strategy also induces stability risks such as control interaction with PE devices, so-called sub-synchronous control interaction \cite{Lingling_SSCI, Ulas_SSCI}.
This instability occurs when the damping of the network resonance mode associated with the series-compensation is smaller than the negative damping introduced by the PE controls, which are then said to excite the subsynchronous oscillation.
The studied case's resonance modes are identified depending on its compensation level $\eta$, and its results are shown in \cref{fig: Freq response compensated branchh}.
Indeed, the resonance mode of compensated lines moves to a lower frequency region as the compensation level increases.

\begin{figure}[!t]
\centering
\includegraphics[width=3.5in]{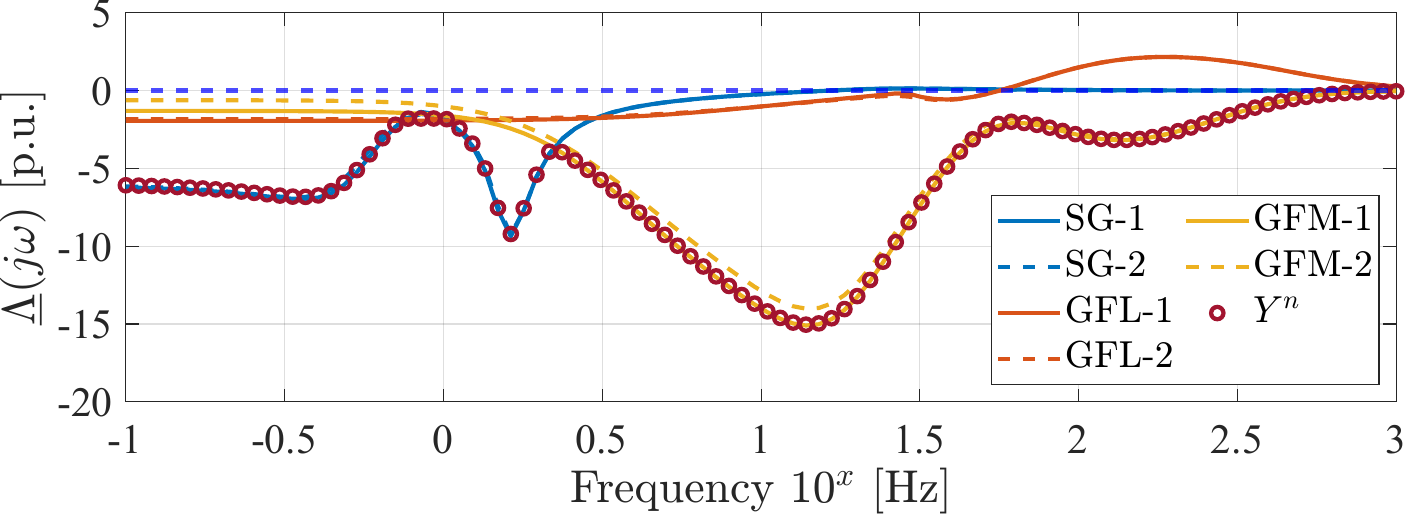}
\caption{Passivity investigation of the two-area system before passivation.}
\label{fig: 2 area NP}
\vspace{-0.3cm}
\end{figure}

\begin{figure}[!t]
\centering
\includegraphics[width=3.5in]{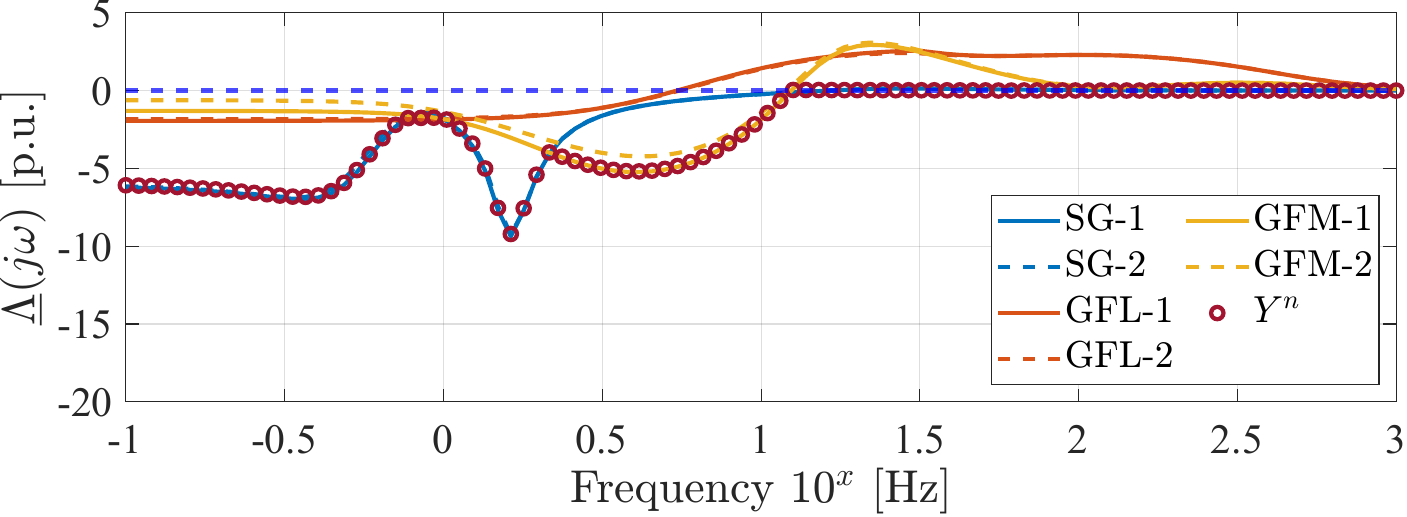}
\caption{Passivity investigation of the two-area system after passivation.}
\label{fig: 2 area P}
\vspace{-0.3cm}
\end{figure}

\begin{figure}[!t]
\centering
\includegraphics[width=3.5in]{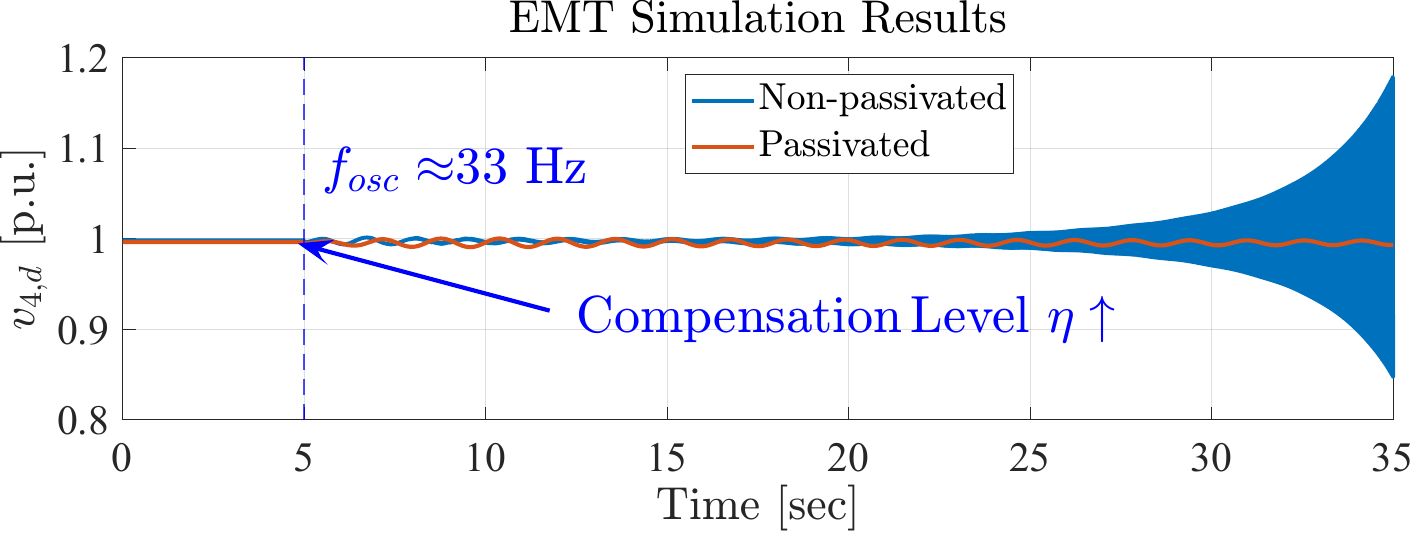}
\caption{EMT simulations of the modified two-area system.}
\label{fig: 2 area EMT}
\vspace{-0.5cm}
\end{figure}

As the network composed of RLC-elements is passive, the passivity approach is especially useful to prevent PE controls from negatively interacting with the network and other devices, and thus achieving a stable integration of PE with large-distance interconnectors.
Firstly, the passivity assessment of the SGs and VSCs subsystems is presented in \cref{fig: 2 area NP}. These results show that the VSCs are the main contributors to the whole system's non-passivity, and that passivity is only guaranteed in the high frequency region around $10^3$~Hz.
Therefore, there is a risk of interaction with the network's resonance mode which falls in the non-passive frequency region as identified from \cref{fig: Freq response compensated branchh}, which can lead to instability.
Since the system topology is hardly changeable compared to device control parameters, the passivation of the VSCs through control parameter modification is used to improve the system passivity properties.
Here, to passivate the GFL VSCs further from previous cases, the voltage feed-forward in the current controller is disabled \cite{Passivity_input_admittance}. 
Passivation for GFM is the same as in the previously introduced case.
Therefore, $\boldsymbol{Y^a}$ is mainly passivated showing $\underline{\Lambda}^n$'s effectiveness on stabilizing the system.
As a result, the passivity assessment for the passivated case is presented in \cref{fig: 2 area P}. 
The results show that the new $\underline{\Lambda}^n$ is highly passivated until $10^{1.1}\approx13$~Hz.
EMT simulations are carried out to confirm the effectiveness of the passivation approach for addressing the identified instability risk. 
The simulation results in \cref{fig: 2 area EMT} show an oscillatory instability around $f_{osc}\approx33$~Hz for the non-passivated case as the compensation level is increased from 20\% to 30\%.
This is expected due to the excitation of the compensated lines' resonance shown in \cref{fig: Freq response compensated branchh} by the non-passivity of $\boldsymbol{Y^a}$ in the same frequency range.
In particular, the network resonance mode shifts from $10^{1.43}\approx26$~Hz to $10^{1.53}\approx34$~Hz as the compensation level increases, and $\underline{\Lambda}^n$ is more negative near the latter system resonance.
Furthermore, the unstable interaction does not occur after passivation in the frequency range of the network resonance.

\section{Conclusion}
This paper developed a passivity sensitivity method for addressing stability challenges with high integration of PE.
The proposed approach is based on the derived passivity sensitivity analysis to extract insights that include not only the magnitude of the interaction but also how different components affect the whole system passivity positively or negatively.
Through the results from the proposed method, the system can be passivated efficiently rather than changing its parameters and checking its impact manually.
Compared to previous sensitivity methods, this is a pure frequency-domain method and it does not require exact system information, such as state-space eigenvalues.

In addition, the proposed method is validated for both device-level parametric sensitivity and system-wide sensitivity.
Each case is shown with application examples: the device-level study shows the effective control design by investigating its parameters' impact on the overall device passivity, and the system-level study shows how systematic sensitivity results point exactly at which devices mostly impact the instability risks at different time-scales.
Both show the proposed method's effectiveness in improving system stability against non-passivated frequency regions.
Finally, a modified two area system is used to show the method's applicability to more complex systems with high penetration of PE, where instability due to PE controls triggering a network resonance mode is successfully mitigated.

The results demonstrate that the proposed method is effective to improve the overall system passivity and stability in a less conservative manner than full passivation of every device.

\appendix[System Parameters]
\subsection{System Parameters}
The parameters used in the validation with case studies are specified in per unit value. Unless otherwise specified, the same parameters as in the previously denoted were used.
System in \cref{fig: 2 area system}: 
System base - $S_b=1000$MVA, $V_b=230$kV, $f_b=60$Hz, 
GFL VSC - $L_c=0.15$, $R_c=0.015$, Current Control - $K_{p,i}=0.75$, $K_{i,i}=37.69$, P\&Q Control - $K_{p,pq}=0.03$, $K_{i,pq}=62.83$, PLL - $K_{p,pll}=0.4$, $K_{i,pll}=30.28$, LPF - $T_i=0.0001$sec, $T_v=0.002$sec, 
GFM VSC - VSM Control - $H_{vsm}=3.0$, $K_{vsm}=10$, $D_{vsm}=300$, Virtual Impedance - $L_v=0.2$, $R_v=0.15$,
SG - $L_{t}=0.15$, $R_{t}=0.015$, $L_{d,a}=1.66$, $L_{d,1}=0.1713$,$L_{d,f}=0.165$, $R_{d,1}=0.02$, $R_{d,f}=0.002$, $L_{q,a}=1.61$, $L_{q,1}=0.7252$, $L_{q,2}=0.125$, $R_{q,1}=0.005$, $R_{q,2}=0.01$, $L_{l}=0.15$, $R_{l}=0.0015$, AVR - $K_A=100$, $T_v=0.05$sec, $T_{1,avr}=1$sec, $T_{2,avr}=0.1$sec, $T_{3,avr}=0.01$sec, Governor - $K_{sg}=10$, $T_{1,gov}=0.1$sec, $T_{2,gov}=0.2$sec, $T_{ch}=2$sec, $T_{rh}=5$sec, $T_{co}=1$sec, $F_{hp}=0.3$, $F_{ip}=0.4$, $F_{lp}=0.3$, Network in \cite{Kundur}, $\eta$=20\%, $P_1, Q_1=0.31, 0.05$, $P_2, V_2=0.45, 1.01$, $P_3, Q_3=-0.95, 0.1$, $P_5, Q_5=-0.95, 0.05$, $P_6, V_6=0.5, 1.01$, $P_7, Q_7=0.65, 0.15$ 
System in \cref{fig: GFL_VSC}: 
System base - $S_b=5$MVA, $V_b=0.6$ kV, $P=0.7$, $Q=0.2$, P\&Q Control - $K_{p,pq}=0.016$, $K_{i,pq}=31.4159$
System in \cref{fig: 3-bus system}: $P_1, Q_1=-0.35, 0.1$, $P_2, Q_2=-0.35, 0.1$, $P_3, Q_3=0.7, 0.0$ 
GFL VSC - $L_c=0.1$, $R_c=0.02$, Current Control - $K_{p,i}=0.5$, 
GFM VSC - $L_c=0.1$, $R_c=0.02$, Current Control - $K_{p,i}=0.5$, $K_{i,i}=37.69$, Network-$Z_{12}^n$=$Z_{13}^n$=$Z_{23}^n=0.086+j0.69$.

\bibliographystyle{IEEEtran}
\bibliography{Main}

\begin{thebibliography}{10}
\providecommand{\url}[1]{#1}
\csname url@samestyle\endcsname
\providecommand{\newblock}{\relax}
\providecommand{\bibinfo}[2]{#2}
\providecommand{\BIBentrySTDinterwordspacing}{\spaceskip=0pt\relax}
\providecommand{\BIBentryALTinterwordstretchfactor}{4}
\providecommand{\BIBentryALTinterwordspacing}{\spaceskip=\fontdimen2\font plus
\BIBentryALTinterwordstretchfactor\fontdimen3\font minus \fontdimen4\font\relax}
\providecommand{\BIBforeignlanguage}[2]{{%
\expandafter\ifx\csname l@#1\endcsname\relax
\typeout{** WARNING: IEEEtran.bst: No hyphenation pattern has been}%
\typeout{** loaded for the language `#1'. Using the pattern for}%
\typeout{** the default language instead.}%
\else
\language=\csname l@#1\endcsname
\fi
#2}}
\providecommand{\BIBdecl}{\relax}
\BIBdecl

\bibitem{IEEEstability2020}
N.~Hatziargyriou, J.~Milanovic, C.~Rahmann, V.~Ajjarapu, C.~Canizares, I.~Erlich, D.~Hill, I.~Hiskens, I.~Kamwa, B.~Pal, P.~Pourbeik, J.~Sanchez-Gasca, A.~Stankovic, T.~Van~Cutsem, V.~Vittal, and C.~Vournas, ``Definition and classification of power system stability – revisited \& extended,'' \emph{IEEE Transactions on Power Systems}, vol.~36, no.~4, pp. 3271--3281, 2021.

\bibitem{HVDCoverview}
N.~R. Watson and J.~D. Watson, ``An overview of hvdc technology,'' \emph{Energies}, vol.~13, no.~17, 2020.

\bibitem{ZvsSS}
M.~Amin and M.~Molinas, ``Small-signal stability assessment of power electronics based power systems: A discussion of impedance- and eigenvalue-based methods,'' \emph{IEEE Transactions on Industry Applications}, vol.~53, no.~5, pp. 5014--5030, 2017.

\bibitem{Sun2022}
J.~Sun, ``Frequency-domain stability criteria for converter-based power systems,'' \emph{IEEE Open Journal of Power Electronics}, vol.~3, pp. 222--254, 2022.

\bibitem{Grey-Box}
Y.~Zhu, Y.~Gu, Y.~Li, and T.~C. Green, ``Participation analysis in impedance models: The grey-box approach for power system stability,'' \emph{IEEE Transactions on Power Systems}, vol.~37, no.~1, pp. 343--353, 2022.

\bibitem{On-Relationship}
C.~Zhang, H.~Zong, X.~Cai, and M.~Molinas, ``On the relation of nodal admittance- and loop gain-model based frequency-domain modal methods for converters-dominated systems,'' \emph{IEEE Transactions on Power Systems}, vol.~38, no.~2, pp. 1779--1782, 2023.

\bibitem{Sainz17}
L.~Sainz, M.~Cheah-Mane, L.~Monjo, J.~Liang, and O.~Gomis-Bellmunt, ``Positive-net-damping stability criterion in grid-connected vsc systems,'' \emph{IEEE Journal of Emerging and Selected Topics in Power Electronics}, vol.~5, no.~4, pp. 1499--1512, 2017.

\bibitem{Cheah17}
M.~Cheah-Mane, L.~Sainz, J.~Liang, N.~Jenkins, and C.~E. Ugalde-Loo, ``Criterion for the electrical resonance stability of offshore wind power plants connected through hvdc links,'' \emph{IEEE Transactions on Power Systems}, vol.~32, no.~6, pp. 4579--4589, 2017.

\bibitem{PMD2021}
L.~Orellana, L.~Sainz, E.~Prieto-Araujo, and O.~Gomis-Bellmunt, ``Stability assessment for multi-infeed grid-connected vscs modeled in the admittance matrix form,'' \emph{IEEE Transactions on Circuits and Systems I: Regular Papers}, vol.~68, no.~9, pp. 3758--3771, 2021.

\bibitem{Orellana2023}
L.~Orellana, L.~Sainz, E.~Prieto-Araujo, M.~Cheah-Mané, H.~Mehrjerdi, and O.~Gomis-Bellmunt, ``Study of black-box models and participation factors for the positive-mode damping stability criterion,'' \emph{International Journal of Electrical Power \& Energy Systems}, vol. 148, p. 108957, 2023.

\bibitem{VF}
B.~Gustavsen and A.~Semlyen, ``Rational approximation of frequency domain responses by vector fitting,'' \emph{IEEE Transactions on Power Delivery}, vol.~14, no.~3, pp. 1052--1061, 1999.

\bibitem{Maciejowski1989}
J.~Maciejowski, \emph{{Multivariable feedback design}}, 1st~ed.\hskip 1em plus 0.5em minus 0.4em\relax Addison-Wesley, 1989.

\bibitem{MTHVDCstability}
Y.~Liao, H.~Wu, X.~Wang, M.~Ndreko, R.~Dimitrovski, and W.~Winter, ``Stability and sensitivity analysis of multi-vendor, multi-terminal hvdc systems,'' \emph{IEEE Open Journal of Power Electronics}, vol.~4, pp. 52--66, 2023.

\bibitem{bao2007}
J.~Bao and P.~Lee, \emph{Process Control: The Passive Systems Approach}, ser. Advances in Industrial Control.\hskip 1em plus 0.5em minus 0.4em\relax Springer London, 2007.

\bibitem{Dey2023}
K.~Dey and A.~M. Kulkarni, ``Passivity-based decentralized criteria for small-signal stability of power systems with converter-interfaced generation,'' \emph{IEEE Transactions on Power Systems}, vol.~38, no.~3, pp. 2820--2833, 2023.

\bibitem{Chatterjee23}
K.~Chatterjee, S.~Samanta, and N.~R. Chaudhuri, ``A passivity-based small-signal def analysis for low-frequency oscillation source characterization of vsc-hvdc,'' \emph{IEEE Transactions on Power Delivery}, vol.~38, no.~6, pp. 4274--4286, 2023.

\bibitem{Wang2024}
D.~Wang and K.~Zhang, ``Small-signal stability analysis of mmc-hvdc system based on hybrid passivity,'' \emph{IEEE Transactions on Power Delivery}, vol.~39, no.~1, pp. 29--41, 2024.

\bibitem{Josep2024}
J.~Arévalo-Soler, M.~Nahalparvari, D.~Groß, E.~Prieto-Araujo, S.~Norrga, and O.~Gomis-Bellmunt, ``Small-signal stability and hardware validation of dual-port grid-forming interconnecting power converters in hybrid ac/dc grids,'' \emph{IEEE Journal of Emerging and Selected Topics in Power Electronics}, pp. 1--1, 2024.

\bibitem{Dey2021}
K.~Dey and A.~Kulkarni, ``Analysis of the passivity characteristics of synchronous generators and converter-interfaced systems for grid interaction studies,'' \emph{International Journal of Electrical Power \& Energy Systems}, vol. 129, p. 106818, 2021.

\bibitem{Feifan2024}
F.~Chen, X.~Wang, L.~Harnefors, S.~Z. Khong, D.~Wang, L.~Zhao, K.~C. Sou, M.~Routimo, J.~Kukkola, H.~Sandberg, and K.~H. Johansson, ``Limitations of using passivity index to analyze grid–inverter interactions,'' \emph{IEEE Transactions on Power Electronics}, vol.~39, no.~11, pp. 14\,465--14\,477, 2024.

\bibitem{Root-Cause}
Y.~Zhu, Y.~Gu, Y.~Li, and T.~C. Green, ``Impedance-based root-cause analysis: Comparative study of impedance models and calculation of eigenvalue sensitivity,'' \emph{IEEE Transactions on Power Systems}, vol.~38, no.~2, pp. 1642--1654, 2023.

\bibitem{Kundur}
P.~Kundur, N.~Balu, and M.~Lauby, \emph{Power System Stability and Control}.\hskip 1em plus 0.5em minus 0.4em\relax McGraw-Hill, 1994.

\bibitem{Bus-PF}
E.~Ebrahimzadeh, F.~Blaabjerg, X.~Wang, and C.~L. Bak, ``Bus participation factor analysis for harmonic instability in power electronics based power systems,'' \emph{IEEE Transactions on Power Electronics}, vol.~33, no.~12, pp. 10\,341--10\,351, 2018.

\bibitem{Harmonic-Resonance}
Z.~Huang, Y.~Cui, and W.~Xu, ``Application of modal sensitivity for power system harmonic resonance analysis,'' \emph{IEEE Transactions on Power Systems}, vol.~22, no.~1, pp. 222--231, 2007.

\bibitem{Oscillatory-Stability}
Y.~Zhan, X.~Xie, H.~Liu, H.~Liu, and Y.~Li, ``Frequency-domain modal analysis of the oscillatory stability of power systems with high-penetration renewables,'' \emph{IEEE Transactions on Sustainable Energy}, vol.~10, no.~3, pp. 1534--1543, 2019.

\bibitem{Passivity_input_admittance}
L.~Harnefors, M.~Bongiorno, and S.~Lundberg, ``Input-admittance calculation and shaping for controlled voltage-source converters,'' \emph{IEEE Transactions on Industrial Electronics}, vol.~54, no.~6, pp. 3323--3334, 2007.

\bibitem{Series_Compensation}
P.~Datka, M.~M. H.~Eriksson, S.~S. K.~Narendra, Z.~B. A.~Taylor, and R.~le~Roux, ``Challenges with series compensation applications in power systems when overcompensating lines,'' \emph{Cigre ELECTRA}, vol. 315, no. TB 829, 2021.

\bibitem{Lingling_SSCI}
Y.~Xu, M.~Zhang, L.~Fan, and Z.~Miao, ``Small-signal stability analysis of type-4 wind in series-compensated networks,'' \emph{IEEE Transactions on Energy Conversion}, vol.~35, no.~1, pp. 529--538, 2020.

\bibitem{Ulas_SSCI}
M.~Ghafouri, U.~Karaagac, J.~Mahseredjian, and H.~Karimi, ``Ssci damping controller design for series-compensated dfig-based wind parks considering implementation challenges,'' \emph{IEEE Transactions on Power Systems}, vol.~34, no.~4, pp. 2644--2653, 2019.

\end{thebibliography}

\vfill
\end{document}